\documentclass{aa}
\usepackage{graphicx}
\usepackage{color}
\usepackage{graphicx}
\usepackage{lscape}
\usepackage{natbib}
\usepackage[varg]{txfonts}
\usepackage{url}
\usepackage{xspace}
\bibpunct{(}{)}{;}{a}{}{,}
\newcounter{Rco}
\newcommand{\Ionst}[1]{\setcounter{Rco}{#1}\Roman{Rco}}
\newcommand{\Ion}[2]{\mbox{#1\,{\scriptsize\Ionst{#2}}}}
\newcommand{\Ionw}[3]{\mbox{#1\,{\scriptsize\Ionst{#2}}~$\lambda\,#3$\,\AA}}

\newcommand{\Ionww}[3]{\mbox{#1\,{\scriptsize\Ionst{#2}}~$\lambda\lambda\,#3$\,\AA}}

\newcommand{\logg}{\mbox{$\log g$}\xspace}
\newcommand{\loggw}[1]{\mbox{$\log g\hspace{-0.5mm} =\hspace{-0.5mm}  #1$}}

\newcommand{\sga}{\raisebox{-0.10em}{$\stackrel{>}{{\mbox{\tiny $\sim$}}}$}}

\newcommand{\Teff}{\mbox{$T_\mathrm{eff}$}\xspace}
\newcommand{\Teffw}[1]{\mbox{$\Teff\hspace{-0.5mm} =\hspace{-0.5mm} #1 \,\mathrm{K}$}}
\newcommand{\ebv}{\mbox{$E_{B-V}$}}

\newcommand{\Msol}{$M_\odot$}

\newcommand{\fg}{\object{Feige\,110}\xspace}
\newcommand{\gb}{\object{G191$-$B2B}\xspace}
\begin{document}
\title{The virtual observatory service TheoSSA: \\
       Establishing a database of synthetic stellar flux standards
      }
\subtitle{II. NLTE spectral analysis of the OB-type subdwarf \fg
           \thanks
           {Based on observations with the NASA/ESA Hubble Space Telescope, obtained at the Space Telescope Science 
            Institute, which is operated by the Association of Universities for Research in Astronomy, Inc., under 
            NASA contract NAS5-26666.
           }$^,$\thanks
           {Based on observations made with the NASA-CNES-CSA Far Ultraviolet Spectroscopic Explorer.
           }
         }

\author{T\@. Rauch\inst{1} 
        \and 
        A\@. Rudkowski\inst{1}
        \and 
        D\@. Kampka\inst{1}
        \and 
        K\@. Werner\inst{1}
        \and 
        J\@. W\@. Kruk\inst{2}
        \and
        S\@. Moehler\inst{3}
        }
 
\institute{Institute for Astronomy and Astrophysics,
           Kepler Center for Astro and Particle Physics,
           Eberhard Karls University, 
           Sand 1,\\
           72076 T\"ubingen,
           Germany\\
           \email{rauch@astro.uni-tuebingen.de}
           \and
           NASA Goddard Space Flight Center, Greenbelt, MD\,20771, USA
           \and
           European Southern Observatory, Karl-Schwarzschild-Str\@. 2, 85748 Garching, Germany
           }

\date{Received 26 February 2014 / Accepted 9 April 2014}

\titlerunning{TheoSSA: Establishing a database of synthetic stellar flux standards. II. \fg}

\abstract {
           In the framework of the Virtual Observatory (VO), the German Astrophysical Virtual Observatory (GAVO)
           developed the registered service TheoSSA (Theoretical Stellar Spectra Access).
           It provides easy access to stellar spectral energy distributions (SEDs) and is
           intended to ingest SEDs calculated by any model-atmosphere code, generally for
           all effective temperature, surface gravities, and elemental compositions.
           We will establish a database of SEDs of flux standards that are easily accessible 
           via TheoSSA's web interface.
          }
          {The OB-type subdwarf \fg is a standard star for flux calibration.
           State-of-the-art 
           non-local thermodynamic equilibrium (NLTE)
           stellar-atmosphere models 
           that consider opacities of species up to trans-iron elements 
           will be used to provide a reliable synthetic spectrum 
           to compare with observations. 
          }
          {
           In case of \fg, we demonstrate that the model reproduces not only its overall continuum shape 
           from the far-ultraviolet (FUV) to the optical wavelength range
           but also the numerous metal lines exhibited in its FUV spectrum. 
          }
          {
           We present a state-of-the-art spectral analysis of \fg.
           We determined 
           \Teffw{47\,250 \pm 2000},
           \loggw{6.00 \pm 0.20}, and
           the abundances of 
           He, N, P, S, Ti, V, Cr, Mn, Fe, Co, Ni, Zn, and Ge.
           Ti, V, Mn, Co, Zn, and Ge were identified for the first time in this star.
           Upper abundance limits were derived for C, O, Si, Ca, and Sc.            
          }
          {
           The TheoSSA database of theoretical SEDs of stellar flux standards 
           guarantees that the flux calibration of astronomical data 
           and cross-calibration between different instruments can be based on
           models and SEDs calculated with state-of-the-art model-atmosphere codes. 
          }
         
\keywords{Standards --
          Stars: abundances -- 
          Stars: atmospheres -- 
          Stars: individual: \fg\ --
          Stars: subdwarfs --
          Virtual observatory tools
         }

\maketitle

\section{Introduction}
\label{sect:intro}

\fg is a bright \citep[$m_\mathrm{V} = 11.845 \pm 0.010$,][]{kharchenkoroeser2009}, subluminous OB-star 
(type sdOB, \citealt{heberetal1984b}; type sdO D,\citealt{vennesetal2011}).
It is widely used as a spectrophotometric standard
star \citep[e.g\@.][]{oke1990,turnsheketal1990,bohlinetal1990}. Since \fg will be used
as a reference star for the flux calibration of 
X-SHOOTER\footnote{\url{http://www.eso.org/sci/facilities/paranal/instruments/xshooter.html}} 
\citep{vernetetal2011} observations from 3000\,\AA\ to 25\,000\,\AA\
\citep{moehleretal2014}, we decided to reanalyze its spectrum with state-of-the-art
model-atmosphere techniques. 

An early spectral analysis with approximate LTE\footnote{local thermodynamic equilibrium},
line-blanketed hydrogen model atmospheres yielded an effective temperature 
\Teffw{39\,000} and a surface gravity $\log (g\,/\,\mathrm{cm/s^2}) = 6.5$ \citep{newell1973}.
\citet{kudritzki1976} showed that both, the consideration of deviations from
LTE (NLTE\footnote{non-local thermodynamic equilibrium}) 
as well as of opacities of elements heavier than H, have a significant
influence on the determination of \Teff and \logg in an analysis of optical
spectra (Table\,\ref{tab:kud76}).
\citet{heberetal1984b} extended the analysis of \fg to the ultraviolet 
(UV) wavelength range (IUE\footnote{International Ultraviolet Explorer} observations,
$1150\,\mathrm{\AA}\,\la\,\lambda\,\la\,2000\,\mathrm{\AA}$) in addition to high-resolution optical spectra 
($4000\,\mathrm{\AA}\,\la\,\lambda\,\la\,5100\,\mathrm{\AA}$) and derived 
\Teffw{40\,000^{+5000}_{-3000}}, 
\loggw{5.0 \pm 0.3}, and
He/H = 0.03$^{+0.03}_{-0.02}$ (by number) using H+He (with subsequent C+N+Si line-formation calculations) NLTE models.

\onltab{
\begin{table}\centering
\caption{\Teff and \logg of \fg determined by \citet{kudritzki1976}.
         He/H gives his models' abundance ratio by number.}
\label{tab:kud76}
\begin{tabular}{rr@{.}lcrr@{.}lr@{.}l}
\hline
\multicolumn{3}{c}{LTE} & & \multicolumn{3}{c}{NLTE}                & \multicolumn{2}{c}{}     \\
\cline{1-3}
\cline{5-7}
\multicolumn{8}{c}{~} \vspace{-5.5mm} \\
\multicolumn{7}{c}{~}                                                & \multicolumn{2}{c}{He/H} \vspace{-2.5mm} \\
\noalign{\smallskip}
\multicolumn{1}{c}{$T_\mathrm{eff}$\,/\,K} & \multicolumn{2}{c}{$\log\,(g\,/\,\mathrm{cm/s^2)}$} & & 
\multicolumn{1}{c}{$T_\mathrm{eff}$\,/\,K} & \multicolumn{2}{c}{$\log\,(g\,/\,\mathrm{cm/s^2)}$} & \multicolumn{2}{c}{}     \\
\noalign{\smallskip}
\hline
\noalign{\smallskip}42\,600 & \hbox{}\hspace{2mm}6&3 && 44\,600 & \hbox{}\hspace{2mm}5&9 & \hbox{}\hspace{2mm}0&1 \\
42\,400 & 6&5 && 42\,700 & 6&4 & 1&0 \\
\hline
\end{tabular}
\end{table}
}

With the FUSE\footnote{Far Ultraviolet Spectroscopic Explorer} mission, the interstellar
deuterium and oxygen column densities toward \fg were measured. \citet{friedmanetal2002}
used optical spectra and estimated the atmospheric parameters by comparison with
a grid of synthetic NLTE model-atmosphere spectra 
\citep[using TLUSTY to compute the stellar atmosphere model and SYNSPEC 
to generate the SED,][just ``TLUSTY'' here after]{hubenylanz1995},
that 
considered 
H and He.
They achieved
\Teffw{42\,300 \pm 1000}, 
\loggw{5.95 \pm 0.15}, and
He/H = $0.011 \pm 0.005$.
With the higher \logg \citep[in agreement with][]{kudritzki1976}, their spectroscopic distance of $d =288 \pm 43\,\mathrm{pc}$ 
agreed with the Hipparcos\footnote{\url{http://www.rssd.esa.int/index.php?project=HIPPARCOS}} 
parallax distance of $d =179^{+265}_{-67}\,\mathrm{pc}$.

In the following, we describe our analysis in detail. In Sect\@.\,\ref{sect:obs}, we
give some remarks on the observations. Then, we introduce our models and the considered
atomic data (Sect\@.\,\ref{sect:models}) and start with a preliminary analysis
(Sect\@.\,\ref{sect:prelim}) of the optical spectrum based on H+He models followed 
by a highly sophisticated analysis with metal-line blanketed models (Sect\@.\,\ref{sect:analysis}).
We summarize our results and conclude in Sect\@.\,\ref{sect:results}.

\section{Observations}
\label{sect:obs}

Our main optical spectrum 
is a median of 19 X-SHOOTER observations,
taken between Oct 26 2011 and July 5 2012 with a $5\arcsec$ slit (the seeing was below $1\arcsec$ during
the observations) and an exposure time of 120\,s each. The achieved resolving power was
$R = \lambda/\Delta\lambda \approx 4800$.
All spectra were extracted with ESO's
standard pipeline-reduction software (with the actual version at the time of the respective observation). 
Heliocentric correction and correction to an airmass = 0 were applied. 
In addition, we used optical HST/STIS\footnote{Hubble Space Telescope / Space Telescope Imaging Spectrograph} spectra 
(ObsIds O40801010 and O40801030 co-added) from the archive for the determination of the interstellar reddening.

Our far-ultraviolet spectrum consists of
two observations of \fg that were performed by FUSE, both in June 2000 and both through the LWRS 
spectrograph aperture.  The dataset IDs were M1080801 and P1044301, with exposure times of 6.2\,ks 
and 21.8\,ks, respectively. Alignment of the four FUSE telescope channels was excellent throughout 
both observations, with RMS exposure-to-exposure variations in flux under 0.5\,\% in all channels. 
The processing of individual exposures to produce a combined spectrum spanning $905-1188\,\mathrm{\AA}$ 
was the same as that described for \gb in \citet{rauchetal2013} and won't be repeated here. The 
signal to noise per 0.013\,\AA\ pixel in the continuum for the combined spectrum is typically 80:1 
shortward of 1000\,\AA\ and 120:1 longward of 1000\,\AA. Approximately 37\,\% of the exposure time 
was obtained during orbital night. Comparison of the spectra obtained during day and night portions 
of the orbit found a discernible difference only in the cores of Ly\,$\beta$ and Ly\,$\gamma$. Only 
the night data were used for these spectral regions. Weak airglow emission was still present at Ly\,$\beta$
during orbital night, but the affected pixels had no impact on the analysis of the stellar spectrum.

Additional ultraviolet spectra were retrieved from MAST. 
We used all available low-resolution IUE
spectra (SWP03737, SWP20091, SWP21888, SWP21890, SWP21891, SWP21892,
LWP01913, LWP01914, LWP01915, LWP02505, LWP02506, LWP02507, LWP02508, and LWR11785 co-added)
and an
HST/STIS spectrum
(ObsId OBIE01010, exposure time 1734.2\,s, start time 2010-12-12 08:10:54 UT, 
grating G140M, $1191\,\mathrm{\AA}\,\la\,\lambda\,\la\,1246\,\mathrm{\AA}$, aperture $52\arcsec \times 0\farcs 05$,
resolution = 0.1\,\AA).

\section{Model atmospheres and atomic data}
\label{sect:models}

For our model-atmosphere calculations, we use the
T\"ubingen NLTE model-atmosphere package\footnote{TMAP, \url{http://astro.uni-tuebingen.de/~TMAP}}
\citep{werneretal2003, rauchdeetjen2003},
that assumes a plane-parallel geometry and considers 
opacities of elements from H to Ni \citep{rauch1997,rauch2003}.
The models are in hydrostatic and radiative equilibrium.
TMAP was successfully used for many spectral analyses of hot, compact stars
\citep[e.g\@.][]{rauchetal2007,wassermannetal2010,klepprauch2011,ziegleretal2012,rauchetal2013}.

The model atoms used in our model-atmosphere calculations were either 
retrieved from the T\"ubingen model-atom database\footnote{TMAD, \url{http://astro.uni-tuebingen.de/~TMAD}} 
or compiled via the registered Virtual Observatory (VO) tool TIRO\footnote{T\"ubingen iron-opacity interface}
that uses Kurucz's atomic data\footnote{\url{http://kurucz.harvard.edu/atoms.html}} 
and line lists \citep[and priv\@. comm.]{kurucz1991, kurucz2009, kurucz2011}. Table\,\ref{tab:statistics}
shows the statistics of our model atoms.

\onltab{
\begin{table*}\centering
\caption{Statistics the atoms used in our calculations. In the case of iron-group
         elements (Ca -- Ni), the super lines include the sample lines
         \citep[Kurucz's LIN lines, cf\@.][]{rauchdeetjen2003}.}         
\label{tab:statistics}
\begin{tabular}{rlrrrrlrrrr}
\hline
\hline
\noalign{\smallskip}
            && \multicolumn{2}{c}{levels} & &             && \multicolumn{2}{c}{levels} & \\
\cline{3-4}
\cline{8-9}
\noalign{\smallskip}
\multicolumn{2}{c}{ion} & NLTE & LTE & lines  & \multicolumn{2}{c}{\hbox{}\hspace{10mm}ion} & NLTE & LTE & super lines & sample lines \\
\hline         
\noalign{\smallskip}
H  & \Ion{}{1} &  15 &   1 &  105 & \hbox{}\hspace{10mm}Ca & \Ion{}{2} &   
                                                       7 &   0 &   26 &       2\,612  \\
   & \Ion{}{2} &   1 &   0 &  $-$ &    & \Ion{}{3} &   7 &   0 &   28 &      40\,664  \\
He & \Ion{}{1} &  29 &  74 &   69 &    & \Ion{}{4} &   7 &   0 &   22 &      20\,291  \\
   & \Ion{}{2} &  20 &  12 &  190 &    & \Ion{}{5} &   7 &   0 &   26 &     141\,956  \\
   & \Ion{}{3} &   1 &   0 &  $-$ &    & \Ion{}{6} &   7 &   0 &   26 &     114\,545  \\
C  & \Ion{}{2} &  16 &  30 &   37 &    & \Ion{}{7} &   1 &   0 &    0 &               \\
   & \Ion{}{3} &  13 &  54 &   32 & Sc & \Ion{}{2} &   7 &   0 &   26 &      77\,014  \\
   & \Ion{}{4} &  54 &   4 &  295 &    & \Ion{}{3} &   7 &   0 &   27 &          687  \\
   & \Ion{}{5} &   1 &   0 &    0 &    & \Ion{}{4} &   7 &   0 &   26 &      15\,024  \\
N  & \Ion{}{2} &  15 & 232 &   18 &    & \Ion{}{5} &   7 &   0 &   24 &     261\,235  \\
   & \Ion{}{3} &  34 &  32 &  129 &    & \Ion{}{6} &   7 &   0 &   26 &     237\,271  \\
   & \Ion{}{4} &  16 &  78 &   30 &    & \Ion{}{7} &   1 &   0 &    0 &               \\
   & \Ion{}{5} &  54 &   8 &  297 & Ti & \Ion{}{2} &   7 &   0 &   27 &     312\,054  \\
   & \Ion{}{6} &   1 &   0 &    0 &    & \Ion{}{3} &   7 &   0 &   25 &      46\,707  \\
O  & \Ion{}{2} &  16 &  31 &   26 &    & \Ion{}{4} &   7 &   0 &   27 &       1\,000  \\
   & \Ion{}{3} &  54 &  18 &  222 &    & \Ion{}{5} &   7 &   0 &   26 &      26\,654  \\
   & \Ion{}{4} &  18 &  76 &   39 &    & \Ion{}{6} &   7 &   0 &   26 &      95\,448  \\
   & \Ion{}{5} &  19 & 107 &   40 &    & \Ion{}{7} &   1 &   0 &    0 &               \\
   & \Ion{}{6} &   1 &   0 &    0 & V  & \Ion{}{2} &   7 &   0 &   27 &     734\,478  \\ 
Si & \Ion{}{3} &  17 &  17 &   28 &    & \Ion{}{3} &   7 &   0 &   25 &     460\,038  \\ 
   & \Ion{}{4} &  16 &   7 &   44 &    & \Ion{}{4} &   7 &   0 &   25 &      37\,130  \\
   & \Ion{}{5} &  25 &   0 &   59 &    & \Ion{}{5} &   7 &   0 &   26 &       2\,123  \\
   & \Ion{}{6} &   1 &   0 &    0 &    & \Ion{}{6} &   7 &   0 &   25 &      35\,251  \\
P  & \Ion{}{4} &  15 &  36 &    9 &    & \Ion{}{7} &   1 &   0 &    0 &               \\
   & \Ion{}{5} &  18 &   7 &   12 & Cr & \Ion{}{2} &   7 &   0 &   27 &     728\,080  \\
   & \Ion{}{6} &   1 &   0 &    0 &    & \Ion{}{3} &   7 &   0 &   27 &  1\,421\,382  \\
S  & \Ion{}{3} &  21 & 210 &   35 &    & \Ion{}{4} &   7 &   0 &   24 &     234\,170  \\
   & \Ion{}{4} &  17 &  83 &   32 &    & \Ion{}{5} &   7 &   0 &   26 &      43\,860  \\
   & \Ion{}{5} &  39 &  71 &  107 &    & \Ion{}{6} &   7 &   0 &   23 &       4\,406  \\
   & \Ion{}{6} &  12 &  25 &   25 &    & \Ion{}{7} &   1 &   0 &    0 &               \\
   & \Ion{}{7} &   1 &   0 &    0 & Mn & \Ion{}{2} &   7 &   0 &   27 &     136\,814  \\
Zn & \Ion{}{3} &   1 &  12 &    0 &    & \Ion{}{3} &   7 &   0 &   27 &  1\,668\,146  \\
   & \Ion{}{4} &   1 &  75 &    0 &    & \Ion{}{4} &   7 &   0 &   25 &     719\,387  \\
   & \Ion{}{5} &  94 &  63 &  785 &    & \Ion{}{5} &   7 &   0 &   25 &     285\,376  \\
   & \Ion{}{6} &   1 &   0 &    0 &    & \Ion{}{6} &   7 &   0 &   24 &      70\,116  \\
Ge & \Ion{}{3} &   1 &  15 &    0 &    & \Ion{}{7} &   1 &   0 &    0 &               \\
   & \Ion{}{4} &   8 &   1 &    8 & Fe & \Ion{}{2} &   7 &   0 &   27 &     531\,170  \\
   & \Ion{}{5} &  29 &  56 &  119 &    & \Ion{}{3} &   7 &   0 &   27 &     537\,689  \\
   & \Ion{}{6} &   1 &   0 &    0 &    & \Ion{}{4} &   7 &   0 &   27 &  3\,102\,371  \\
\multicolumn{5}{c}{}              &    & \Ion{}{5} &   7 &   0 &   25 &  3\,266\,247  \\
\multicolumn{5}{c}{}              &    & \Ion{}{6} &   7 &   0 &   22 &     991\,935  \\
\multicolumn{5}{c}{}              &    & \Ion{}{7} &   1 &   0 &    0 &               \\
\multicolumn{5}{c}{}              & Co & \Ion{}{2} &   7 &   0 &   27 &     593\,140  \\
\multicolumn{5}{c}{}              &    & \Ion{}{3} &   7 &   0 &   27 &  1\,325\,205  \\
\multicolumn{5}{c}{}              &    & \Ion{}{4} &   7 &   0 &   27 &     552\,916  \\
\multicolumn{5}{c}{}              &    & \Ion{}{5} &   7 &   0 &   27 &  1\,469\,717  \\
\multicolumn{5}{c}{}              &    & \Ion{}{6} &   7 &   0 &   25 &     898\,484  \\
\multicolumn{5}{c}{}              &    & \Ion{}{7} &   1 &   0 &    0 &               \\
\multicolumn{5}{c}{}              & Ni & \Ion{}{2} &   7 &   0 &   27 &     322\,269  \\
\multicolumn{5}{c}{}              &    & \Ion{}{3} &   7 &   0 &   26 &  1\,033\,920  \\
\multicolumn{5}{c}{}              &    & \Ion{}{4} &   7 &   0 &   27 &  2\,512\,561  \\
\multicolumn{5}{c}{}              &    & \Ion{}{5} &   7 &   0 &   27 &  2\,766\,664  \\
\multicolumn{5}{c}{}              &    & \Ion{}{6} &   7 &   0 &   27 &  7\,408\,657  \\
\multicolumn{5}{c}{}              &    & \Ion{}{7} &   1 &   0 &    0 &              \\
\hline                                                                                     
\multicolumn{5}{r}{total}         & 19 &        93 & 1021& 1435& 3958 & 35\,286\,864 \\
\hline
\end{tabular}
\end{table*}  
}

\section{Preliminary analysis}
\label{sect:prelim}

\begin{figure}
  \resizebox{\hsize}{!}{\includegraphics{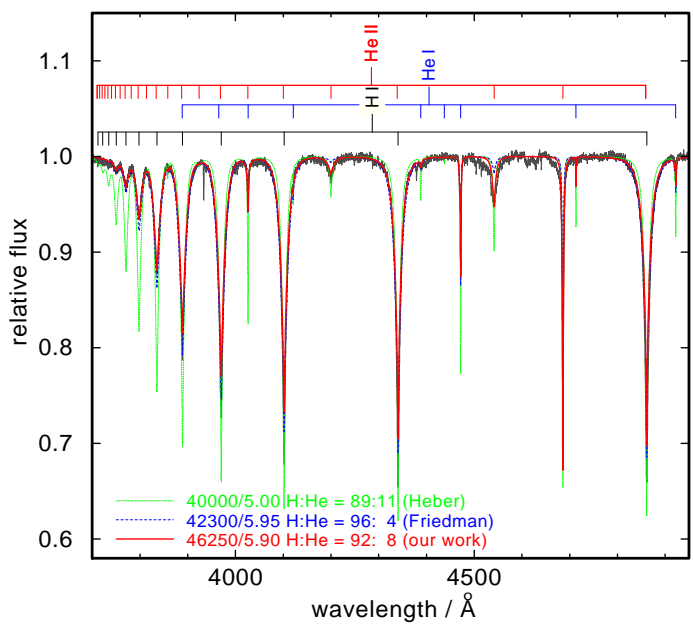}}
  \caption{Comparison of three synthetic spectra with our optical observation of \fg.
           \Teff, \logg, and the H:He ratio by mass are indicated.} 
  \label{fig:optprel}
\end{figure}

For a preliminary analysis, or verification of basic previous results,
we employ the registered Virtual Observatory (VO) service 
TheoSSA\footnote{Theoretical Stellar Spectra Access, \\ \hbox{}\hspace{4mm}\url{ http://dc.g-vo.org/theossa}}
and the related registered VO tool
TMAW\footnote{T\"ubingen Model-Atmosphere WWW Interface}, 
to download pre-calculated synthetic spectral energy distributions (SEDs)
or to calculate individual SEDs, respectively \citep[cf\@.][]{rauchetal2013}.
Figure\,\ref{fig:optprel} shows a comparison of SEDs with model parameters of
\citet{heberetal1984b}, \citet{friedmanetal2002}, and of this work with the observed optical spectrum.
The Balmer decrement is a sensitive indicator of \logg \citep[e.g\@.][]{rauchetal1998},
and we derive \loggw{5.90 \pm 0.20}. At this \logg, the \ion{He}{i} / \ion{He}{ii} ionization
equilibrium, i.e\@. the measured equivalent-width ratio of \ion{He}{i} and \ion{He}{ii} lines,
is well reproduced by our model at \Teffw{46\,250 \pm 2000}. The He line strengths are matched at a
photospheric He abundance of $8 \pm 2$\,\% by mass. 
Although the theoretical H and He line profiles agree well with the observation, the central
depressions are not matched perfectly. This may be a hint of a weak Balmer-line problem 
\citep[cf\@.][]{napiwotzkirauch1994,rauch2000} exists because metal opacities are neglected.
For the same reason, our synthetic H+He SEDs are not suitable for an analysis of the
\ion{H}{i} Lyman lines in FUV\footnote{far-ultraviolet} spectrum. Fully metal-line blanketed
model-atmospheres are mandatory for this purpose (Sect.\,\ref{sect:analysis}). However, 
from our derived \Teff, we are well in the parameter range where no deviation between
\ion{H}{i} Lyman- and Balmer-line analysis is expected \citep[][their Fig.\,4]{goodetal2004}.

\begin{figure}
  \resizebox{\hsize}{!}{\includegraphics{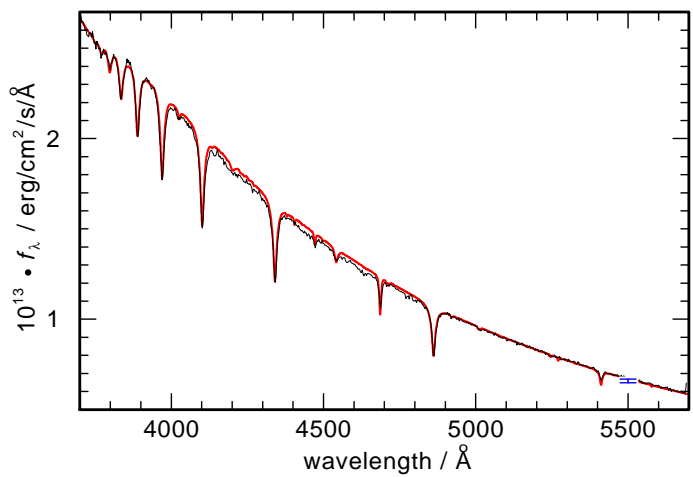}}
  \caption{Comparison of the optical HST/STIS observation with our final model SED.
           The synthetic spectrum is convolved with a Gaussian (FWHM = 5\,\AA) to match
           the resolution of the observation.
           The error bar indicates the visual brightness ($m_\mathrm{V} = 11.847 \pm 0.010$).} 
  \label{fig:opthst}
\end{figure}

We adopt our derived \Teff and \logg values, that also reproduce well the HST/STIS observation
(Fig.\,\ref{fig:opthst}), 
for our further analysis and will verify them with our final, fully metal-line blanketed model.

Within error limits, our preliminary values agree well with those of \citet{friedmanetal2002}.
Only $\Delta T_\mathrm{eff} = \pm 1000$ from their $\chi^2$ fit appears to be too
optimistic. It is worthwhile to note, that the result of \citet{kudritzki1976} (Table\,\ref{tab:kud76},
his lower He abundance model) is relatively close to our result.

\section{Line identification and detailed analysis}
\label{sect:analysis}

\citet{friedmanetal2002} identified photospheric lines from 
\ion{N}{iii - v}, 
\ion{S}{vi}, 
\ion{Cr}{iv - v}, 
\ion{Fe}{iii - iv}, and
\ion{Ni}{iv} in the FUSE observation.
Their SED calculation \citep[SYNSPEC,][]{hubenylanz1995} included all elements from H to Zn, 
all with solar abundances but
He ($1.3 \times 10^{-1}$ times solar),
C ($\approx 4 \times 10^{-6}$ times solar),
Si ($\approx 2 \times 10^{-7}$ times solar),
Cr ($\approx 21$ times solar).

We decided to include 
H, He, C, N, O, Si, P, S, Ca, Sc, Ti, V, Cr, Mn, Fe, Co, Ni, Zn, and Ge
in our calculations. Figure\,\ref{fig:ionfrac} shows the ionization fractions
of these elements. For the iron-group elements (here Ca - Ni), the dominant ionization
stages are {\sc iv - v}.  All SEDs that were calculated for this analysis are available via
TheoSSA.

\onlfig{
\begin{figure*}
  \resizebox{\hsize}{!}{\includegraphics{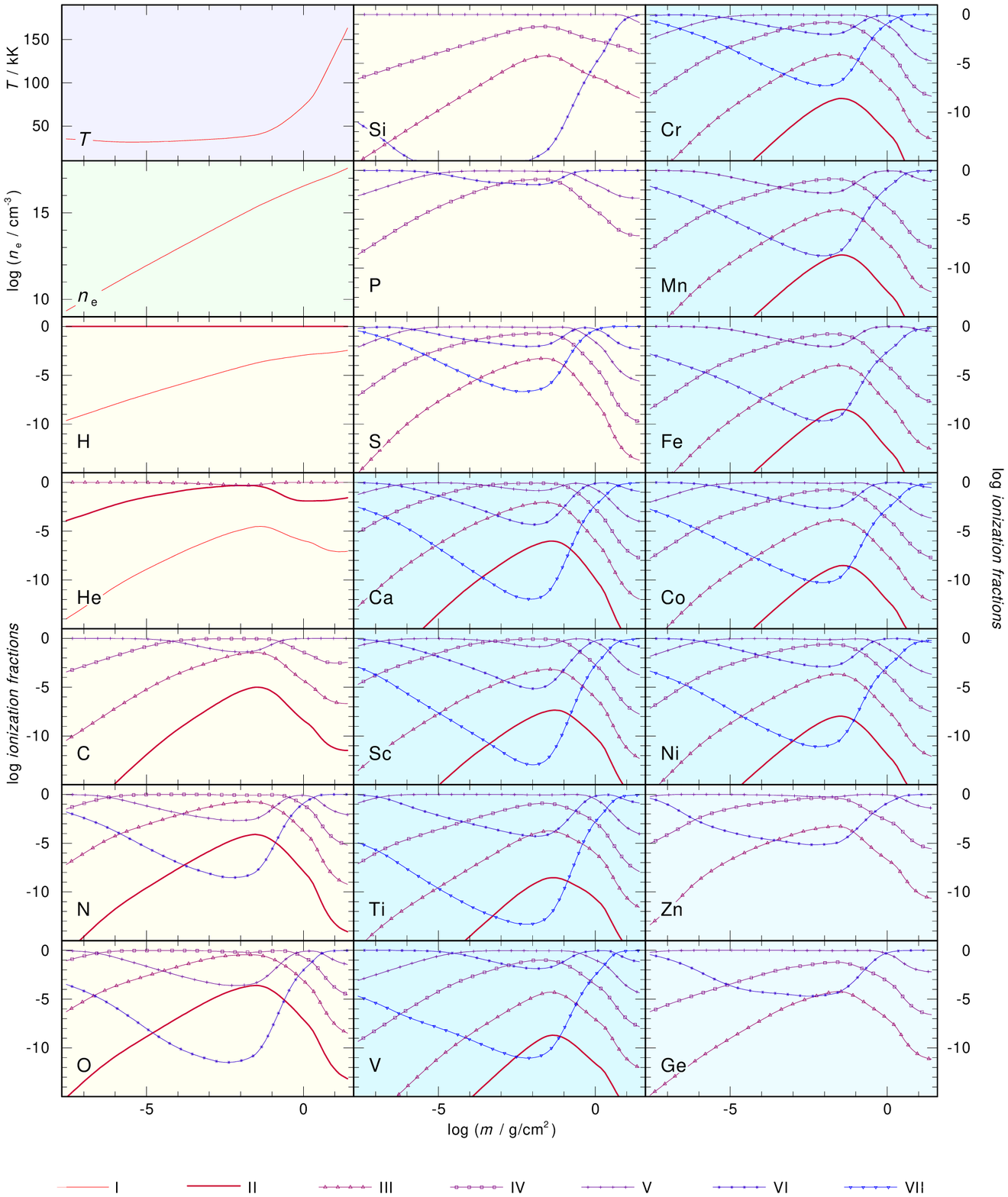}}
  \caption{Temperature and density structure and ionizations fractions of our final model
           with \Teffw{47\,250} and \loggw{6.00}. $m$ is the column mass, measured
           from the outer boundary of our model atmosphere.}
  \label{fig:ionfrac}
\end{figure*}
}

For the line identification in the FUSE wavelength range ($905 - 1188\,\mathrm{\AA}$),
it was necessary to determine the stellar continuum flux precisely. We started with a
measurement of the interstellar neutral hydrogen column density. From
Ly\,$\alpha$ in the STIS spectrum and the higher members of the 
Lyman series in the FUSE spectrum, we determined 
$n_\ion{H}{i} = 1.8 \pm 0.8 \times 10^{20}\,\mathrm{cm^{-2}}$ in agreement with
\citet[$n_\ion{H}{i} = 1.4 \pm 0.5 \times 10^{20}\,\mathrm{cm^{-2}}$ determined
from the high-resolution IUE spectrum SWP15270]{friedmanetal2002}.
To measure the interstellar reddening, we normalized our
synthetic spectrum to the 2MASS H brightness because the interstellar reddening is
negligible there and adjusted
$\ebv$ to match the IUE, STIS, and FUSE flux levels. Our result is
$\ebv = 0.027 \pm 0.007$. 
This very small value is in agreement with the absence of the 2175\,\AA\ bump in the
IUE LWP spectra (Sect.\,\ref{sect:obs}).
The Galactic reddening law of 
\citet[][valid for $0.015\,\la\,E_{B-V}\,\la\,0.075$ and $|b| \ge 20\degr$]{liszt2014a,liszt2014b}, 
$N_\ion{H}{i}/E_{B-V} = 8.3 \times 10^{21}\,\mathrm{cm^{-2}mag^{-1}}$, 
predicts $0.012\,\le\,E_{B-V}\,\le\,0.031$ in agreement with our value. 

\begin{figure}
  \resizebox{\hsize}{!}{\includegraphics{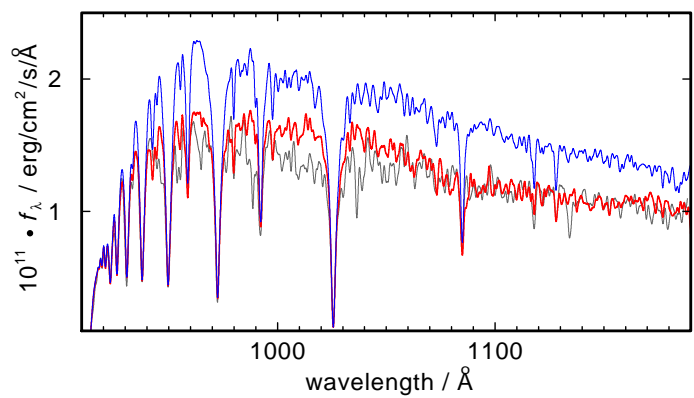}}
  \caption{FUSE observation of \fg (gray) compared with two synthetic spectra calculated
           from our final model 
           (thin, blue in the online version: with Kurucz's POS lines;
            thick, red: with Kurucz's LIN lines).
           All spectra are convolved with a Gaussian (FWHM = 1\,\AA) for clarity.} 
  \label{fig:fusecont}
\end{figure}

\begin{figure}
  \resizebox{\hsize}{!}{\includegraphics{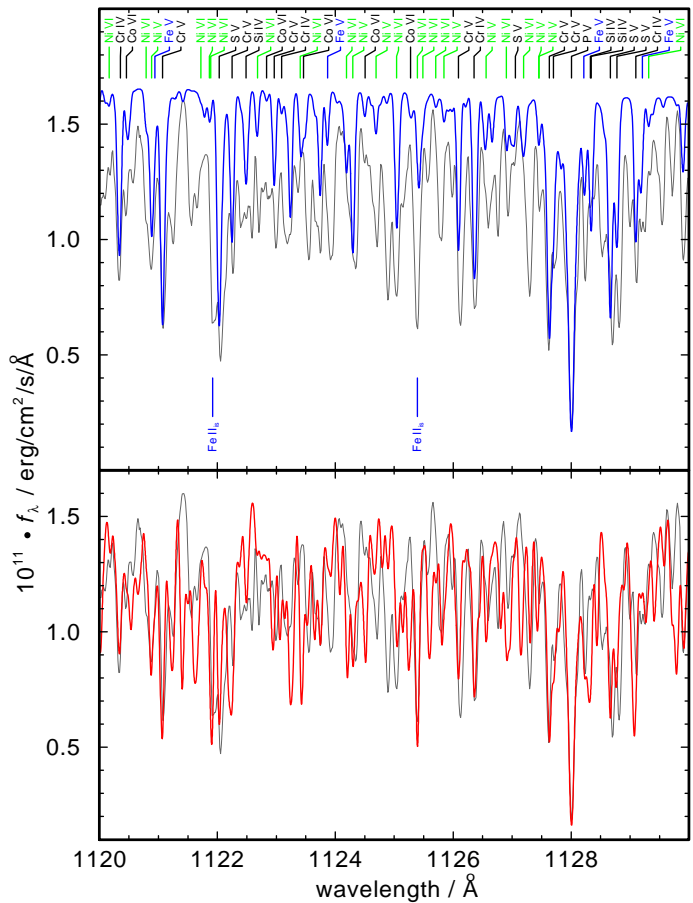}}
  \caption{Same as Fig.\,\ref{fig:fusecont}, for a section of the FUSE observation
           (top panel: POS, bottom panel: LIN). 
            The POS lines are identified 
            (in black, Fe and Ni lines in blue and green for clarity, respectively) 
            at top of
            the top panel.
            Lines of interstellar origin are marked in blue (with subscript ``is'')
            at the bottom of the top panel.
            The synthetic spectra a convolved with a Gaussian (FWHM = 0.06\,\AA) to
            match FUSE's resolution.} 
  \label{fig:friedman}
\end{figure}

The comparison of our models to the FUSE observations shows that we can reproduce well
the observed flux level (Fig.\,\ref{fig:fusecont}), if we include all the lines from Kurucz's
LIN lists (Sect.\,\ref{sect:models}). These include laboratory-measured lines with 
``good wavelengths'' as well as theoretical lines\footnote{Kurucz's LIN lists are
used in our model-atmosphere calculations.}. The lines with good wavelengths are
presented in Kurucz's POS lists. Unfortunately, the ratio of LIN to POS lines is
about 100 and thus, most line wavelengths are uncertain. 
Moreover, the continuum flux of the POS-line spectrum appears artificially high 
compared to the LIN spectrum due to the neglected line opacity 
(Fig.\,\ref{fig:friedman}).

The line-identification process is easy (the comparison of two SEDs calculated from
our final model where the oscillator strengths of one individual atom/ion was artificially
reduced for one SED). It enabled us to unambiguously identify hundreds of lines of
N, O, P, S, Ti, V, Cr, Mn, Fe, Mn, Ni, Zn, and Ge.

\begin{figure}
  \resizebox{\hsize}{!}{\includegraphics{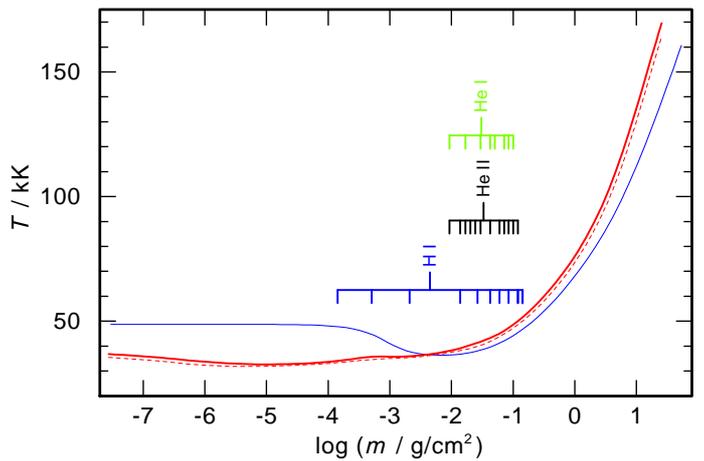}}
  \caption{Temperature structure of our H+He model (thin, blue line: \Teffw{46\,250}, \loggw{5.90}),
           a metal-line blanketed model with the same \Teff and \logg (dashed, red),
           and of our final model (thick, red: \Teffw{47\,250}, \loggw{6.00}).
           The formation depths of the lines cores of optical lines of
           \ion{H}{i} (H\,$\alpha$ is the most outside) , 
           \ion{He}{i},
           and \ion{He}{ii} are shown.} 
  \label{fig:Tstructure}
\end{figure}

Our metal-line blanketed models have a different atmospheric structure compared to the
H-He models that were used in the preliminary determination of \Teffw{46\,250} and \loggw{5.90}
(Sect.\,\ref{sect:prelim}). Figure\,\ref{fig:Tstructure} shows the typical surface-cooling
($\log m\,\la\,\mbox{-2.5}$) and backwarming effects ($\log m\,\sga\,-2.5$), that are an impact
of the additionally 
considered metal opacities. A detailed evaluation of the optical spectrum shows that
slightly higher \Teffw{47\,250} and \loggw{6.00} values are necessary to reproduce the 
\ion{He}{i}\,/\,\ion{He}{ii} ionization equilibrium and the observed \ion{H}{i}, \ion{He}{i},
and \ion{He}{ii} line profiles best.

The abundance analysis follows a standard procedure. Identified lines are reproduced
by an abundance adjustment of the respective species. For elements with no lines
identified, we increased the abundances until their line-detection limit. 
The optical spectrum was used to further constrain the upper limit because
lines of lower ionization stages, that are not observed, appear there 
at too-high abundances in the synthetic spectrum.
Table\,\ref{tab:abund} summarizes the lines that were used and the derived abundances.

\begin{table*}\centering
\caption{Strategic lines and determined element abundances (mass fraction, error $\pm 0.2$\,dex).
         [X] = log (abundance / solar abundance) of species X \citep[solar values from ][]{asplundetal2009}.}
\label{tab:abund}
\begin{tabular}{rp{10cm}r@{.}lr@{.}l}
\hline
\hline
element & analyzed lines                            & \multicolumn{2}{r}{abundance}   & \multicolumn{2}{r}{[X]}   \\
\hline
\noalign{\smallskip}
     H  & optical \ion{H}{i} lines                  & $  9$&$12 \times 10^{-1}$       & $  0$&$00$                \\
     He & optical \ion{He}{i - ii} lines            & $  7$&$93 \times 10^{-2}$       & $ -0$&$50$                \\
     C  & \Ionww{C}{3}{1174-1176}  \newline                      
          \Ionww{C}{4}{1118.41, 1122.49, 1168-1169}  & $< 1$&$04 \times 10^{-7}$       & $<-4$&$36$                \\
     N  & \Ionww{N}{3}{1182.97, 1183.03, 1184.51, 1184.57} \newline
          \Ionww{N}{3}{3998.63, 4003.58, 4379.11, 4510.91, 4514.86, 4634.14, 4640.64} \newline                                         
          \Ionww{N}{4}{921.99, 922.52,  923.06,  923.22,  923.68,  924.28} \newline                                         
          \Ionww{N}{5}{1238.82, 1242.80}            & $  1$&$56 \times 10^{-4}$       & $ -0$&$65$                \\
     O  & \Ionw{O}{3}{1153.78}  \newline                      
          \Ionww{O}{4}{921.30, 921.36, 923.37, 923.43}                                                                    
                                                    & $< 1$&$78 \times 10^{-6}$       & $<-3$&$51$                \\
     Si & \Ionw{Si}{3}{1113.23}  \newline                      
          \Ionw{Si}{4}{1122.49}                     & $< 3$&$96 \times 10^{-7}$       & $<-3$&$23$                \\
     P  & \Ionww{P}{4}{1025.56, 1028.09, 1030.51, 1030.51, 1033.11, 1035.52}  \newline                      
          \Ionww{P}{5}{1117.98, 1128.01}            & $  6$&$67 \times 10^{-6}$       & $  0$&$06$                \\
     S  & \Ionww{S}{4}{1062.66, 1072.97, 1073.52, 1098.93, 1099.48} \newline                                
          \Ionww{S}{5}{1039.92, 1122.03, 1128.67, 1128.78} \newline                                
          \Ionww{S}{6}{933.38, 944.52}              & $  9$&$77 \times 10^{-5}$       & $ -0$&$50$                \\
     Ca & optical \ion{Ca}{iv} lines                & $< 9$&$27 \times 10^{-5}$       & $< 0$&$16$                \\
     Sc & \Ionw{Sc}{4}{931.42} \newline
          \Ionww{Sc}{5}{939.40, 944.04}             & $< 3$&$08 \times 10^{-4}$       & $< 3$&$82$                \\
     Ti & optical \ion{Ti}{iv} lines, \Ionw{Ti}{4}{1183.63} \newline
          \Ionww{Ti}{5}{1153.28, 1163.52}           & $  1$&$77 \times 10^{-4}$       & $  2$&$24$                \\
     V  & \Ionw{V}{4}{1131.25} \newline
          \Ionww{V}{5}{978.16, 1142.74, 1157.58}    & $  5$&$50 \times 10^{-5}$       & $  2$&$06$                \\
     Cr & many \ion{Cr}{iv - vi} lines in the FUV, e.g\@. \newline
          \Ionww{Cr}{4}{1043.46, 1065.26, 1072.10, 1096.64, 1126.35} \newline
          \Ionww{Cr}{5}{1031.10, 1035.04, 1042.55, 1045.04, 1060.65} \newline
          \Ionw{Cr}{6}{957.01}                      & $  1$&$92 \times 10^{-3}$       & $  2$&$06$                \\
     Mn & \Ionww{Mn}{5}{1040.04, 1043.65, 1048.63, 1049.43, 1055.98, 1062.49, 1172.06}  \newline 
          \Ionww{Mn}{6}{1081.09, 1113.58}           & $  1$&$92 \times 10^{-3}$       & $  2$&$25$                \\
     Fe & many \ion{Fe}{v - vi} lines in the FUV, e.g\@. \newline
          \Ionww{Fe}{5}{1002.87, 1015.33, 1020.36}  \newline 
          \Ionww{Fe}{6}{1000.93, 1167.70}           & $  1$&$08 \times 10^{-3}$       & $ -0$&$08$                \\
     Co & many \ion{Co}{v - vi} lines in the FUV, e.g\@. \newline
          \Ionww{Co}{5}{1179.59, 1183.91, 1184.60}  \newline 
          \Ionww{Co}{6}{1133.71, 1142.77, 1150.23, 1169.55, 1175.36}
                                                    & $  8$&$72 \times 10^{-4}$       & $  2$&$32$                \\
     Ni & many \ion{Ni}{v - vi} lines in the FUV, e.g\@. \newline
          \Ionww{Ni}{5}{1124.30, 1178.92}  \newline 
          \Ionww{Ni}{6}{1000.39, 1157.55, 1159.00, 1178.37} 
                                                    & $  2$&$28 \times 10^{-3}$       & $  1$&$51$                \\    
     Zn & \Ionww{Zn}{5}{1116.84, 1120.33, 1158.76}  & $  9$&$08 \times 10^{-5}$       & $  1$&$72$                \\
     Ge & \Ionww{Ge}{5}{1016.67, 1069.13, 1072.66, 1116.95, 1165.26}
                                                    & $  5$&$38 \times 10^{-5}$       & $  2$&$36$                \\
\hline
\end{tabular}
\end{table*}

To identify ISM\footnote{interstellar medium} absorption lines
in the FUSE observation \citep[cf\@.]{friedmanetal2002} and to
judge the contamination of photospheric lines, 
we follow our standard procedure and model the stellar spectrum simultaneously with the 
ISM line absorption \citep[e.g\@.,][]{rauchetal2013}. We modeled the latter with the program 
OWENS \citep{hebrard02, hebrard03},
that considers different clouds with individual radial and turbulent velocities, 
temperatures, column densities and chemical compositions.
Lines are represented by Voigt profiles. The best fit is determined via a $\chi^2$ method.     
Our ISM model includes lines of 
H$_2$ ($J=0-9$), 
\ion{H}{i}, 
\ion{D}{i}, 
\ion{C}{ii-iii}, 
\ion{N}{i-ii},
\ion{O}{i}, 
\ion{Si}{i-ii}, 
\ion{P}{ii}, 
\ion{Ar}{i}, and
\ion{Fe}{ii}.
Our results for \ion{D}{i} and \ion{O}{i} are consistent with those of \citet{friedmanetal2002}.

The complete FUSE observation is compared (including line identifications) with our final model in 
Fig.\,\ref{fig:FUSEcomplete}.

\onlfig{
\begin{landscape}
\addtolength{\textwidth}{6.3cm} 
\addtolength{\evensidemargin}{1cm}
\addtolength{\oddsidemargin}{1cm}
\begin{figure*}
\includegraphics[trim=0 0 0 0,height=24.5cm,angle=270]{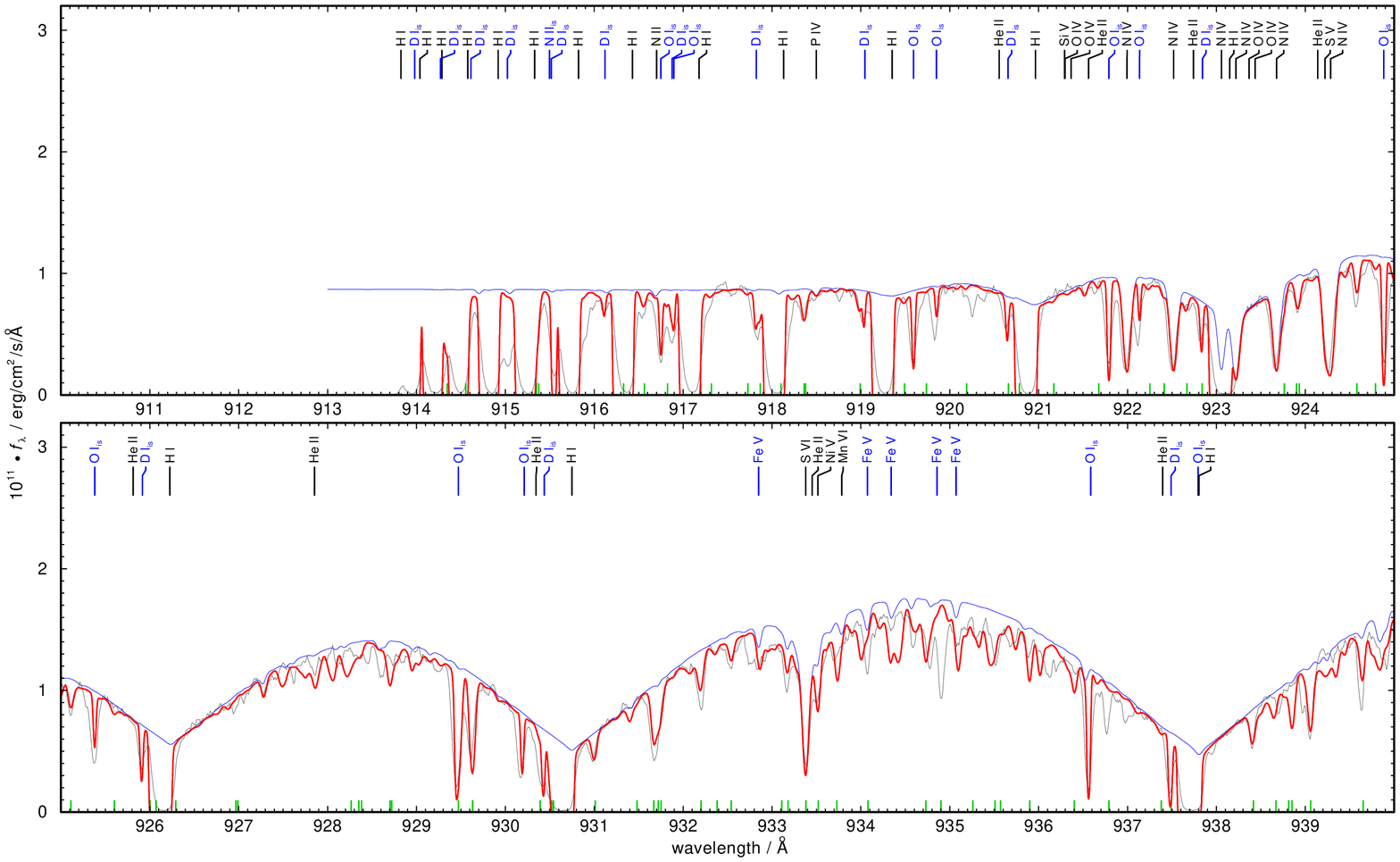}
  \caption{FUSE spectrum of \fg (gray) compared with synthetic spectra calculated from our
           final model
           (red: Kurucz's LIN lines and interstellar line absorption,
           blue: Kurucz's POS lines). 
           The locations of the strongest stellar
           (marked in black, Fe and Ni which have the most lines are marked in blue and green for clarity, respectively,
           for Ca - Ni, only Kurucz's POS lines are identified)
           and 
           interstellar lines 
           (marked in blue, with subscript ``is'')
           are indicated at the top of the panels.
           The small, green identification marks at the bottom  of the panels indicate 
           the locations of interstellar H$_2$ lines.
          }
  \label{fig:FUSEcomplete}
\end{figure*}
\end{landscape}
}

\addtocounter{figure}{-1}

\onlfig{
\begin{landscape}
\addtolength{\textwidth}{6.3cm} 
\addtolength{\evensidemargin}{1cm}
\addtolength{\oddsidemargin}{1cm}
\begin{figure*}
\includegraphics[trim=0 0 0 0,height=24.5cm,angle=270]{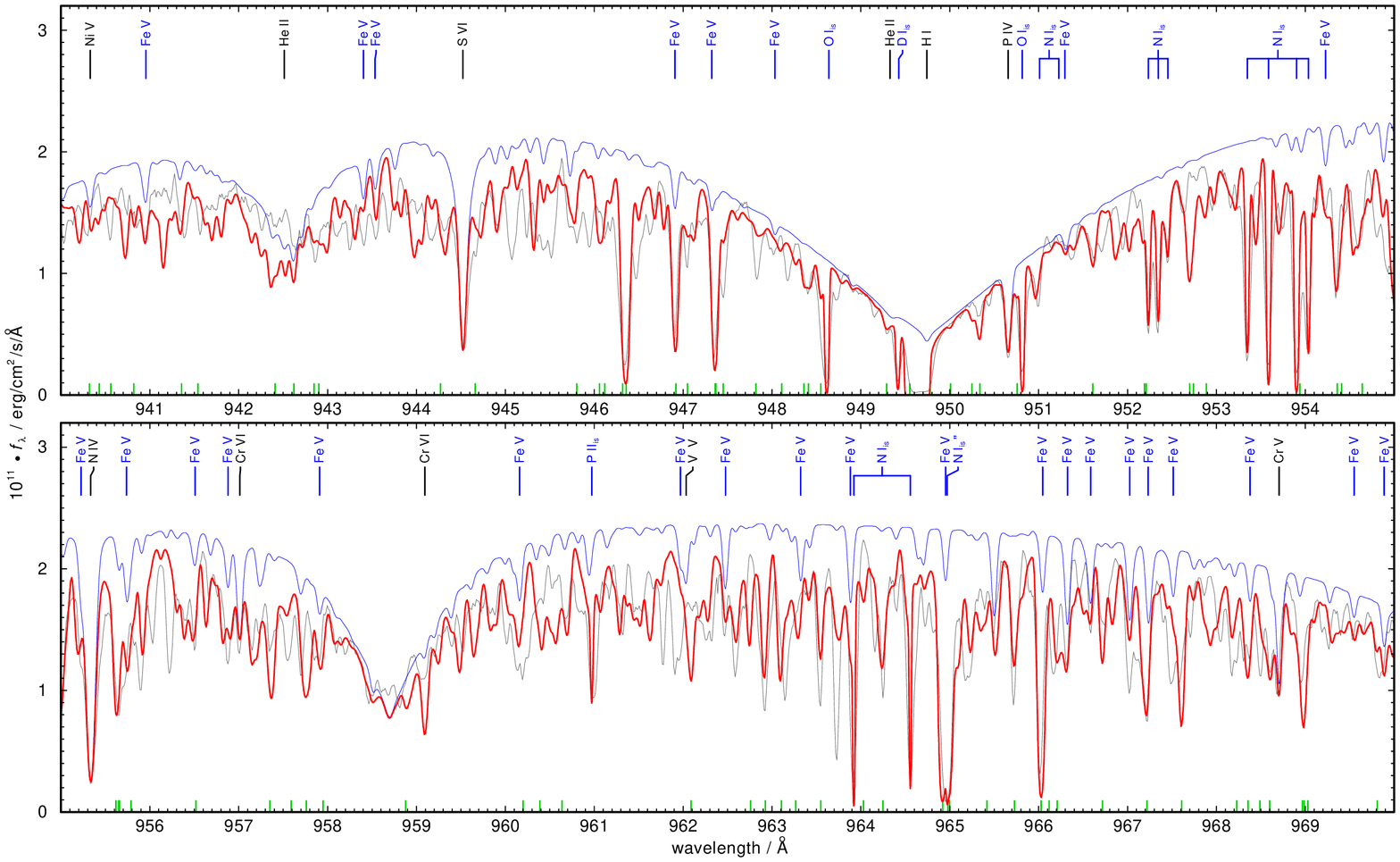}
  \caption{Continued.}
\end{figure*}
\end{landscape}
}

\addtocounter{figure}{-1}

\onlfig{
\begin{landscape}
\addtolength{\textwidth}{6.3cm} 
\addtolength{\evensidemargin}{1cm}
\addtolength{\oddsidemargin}{1cm}
\begin{figure*}
\includegraphics[trim=0 0 0 0,height=24.5cm,angle=270]{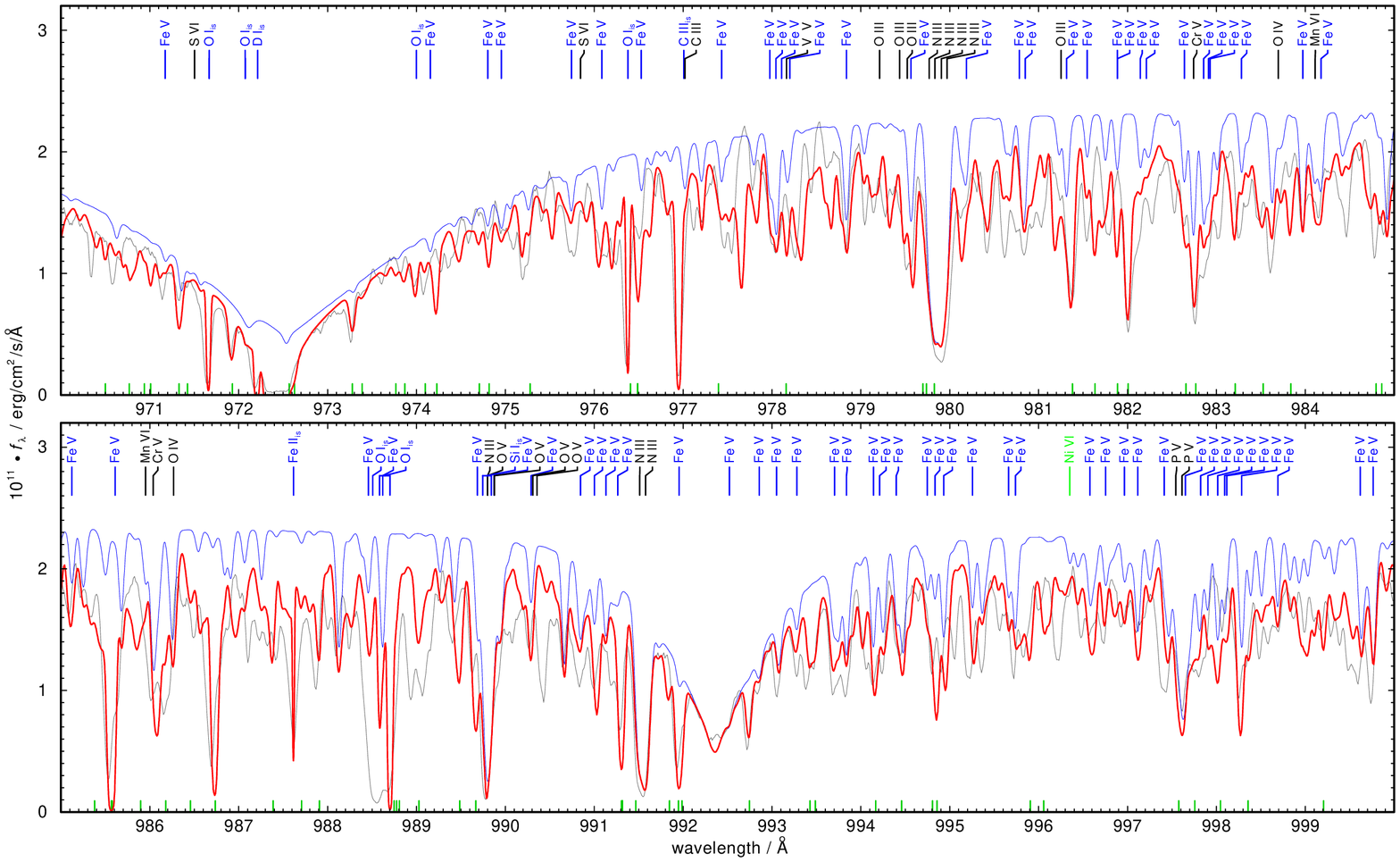}
  \caption{Continued.}
\end{figure*}
\end{landscape}
}

\addtocounter{figure}{-1}

\onlfig{
\begin{landscape}
\addtolength{\textwidth}{6.3cm} 
\addtolength{\evensidemargin}{1cm}
\addtolength{\oddsidemargin}{1cm}
\begin{figure*}
\includegraphics[trim=0 0 0 0,height=24.5cm,angle=270]{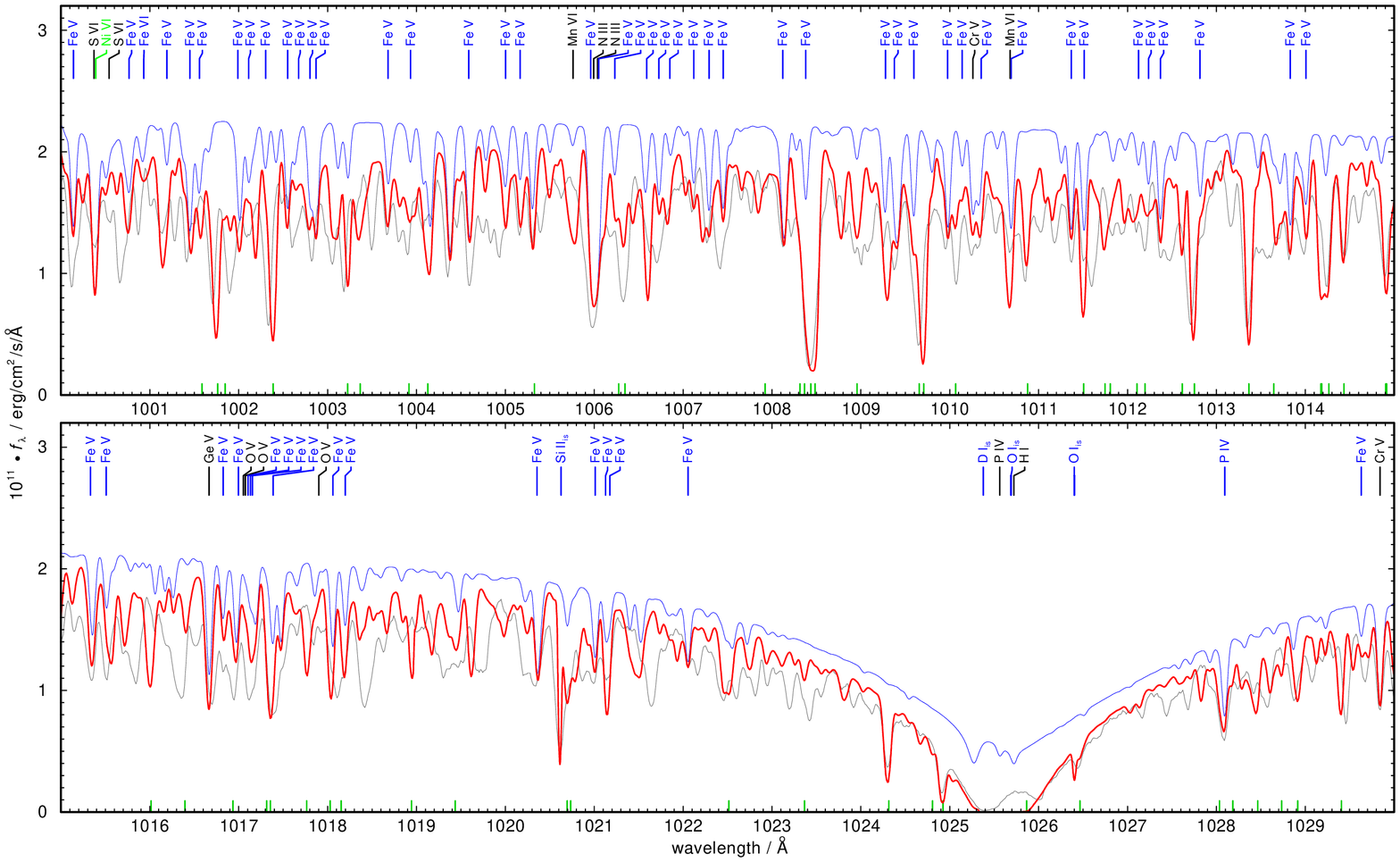}
  \caption{Continued.}
\end{figure*}
\end{landscape}
}

\addtocounter{figure}{-1}

\onlfig{
\begin{landscape}
\addtolength{\textwidth}{6.3cm} 
\addtolength{\evensidemargin}{1cm}
\addtolength{\oddsidemargin}{1cm}
\begin{figure*}
\includegraphics[trim=0 0 0 0,height=24.5cm,angle=270]{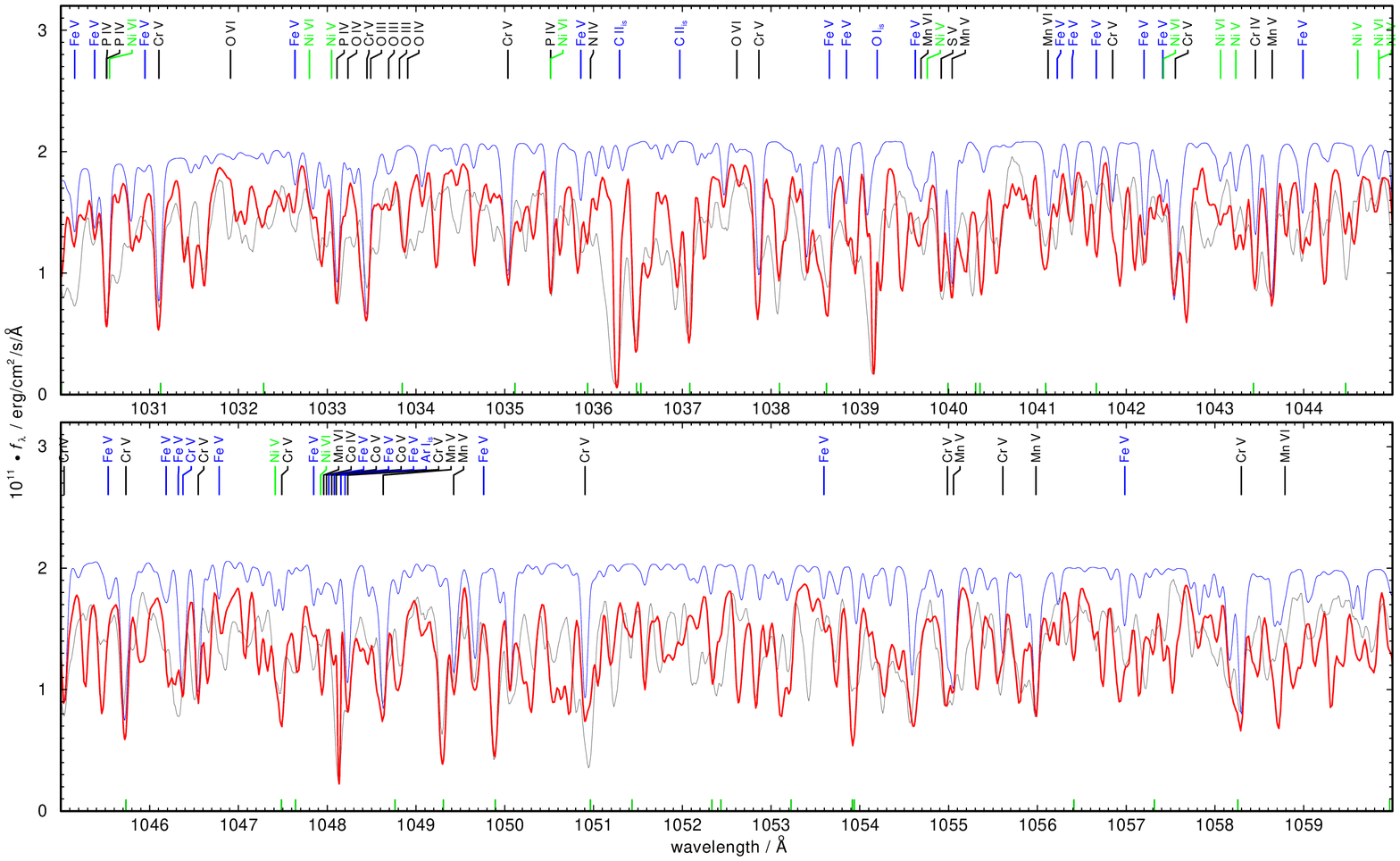}
  \caption{Continued.}
\end{figure*}
\end{landscape}
}

\addtocounter{figure}{-1}

\onlfig{
\begin{landscape}
\addtolength{\textwidth}{6.3cm} 
\addtolength{\evensidemargin}{1cm}
\addtolength{\oddsidemargin}{1cm}
\begin{figure*}
\includegraphics[trim=0 0 0 0,height=24.5cm,angle=270]{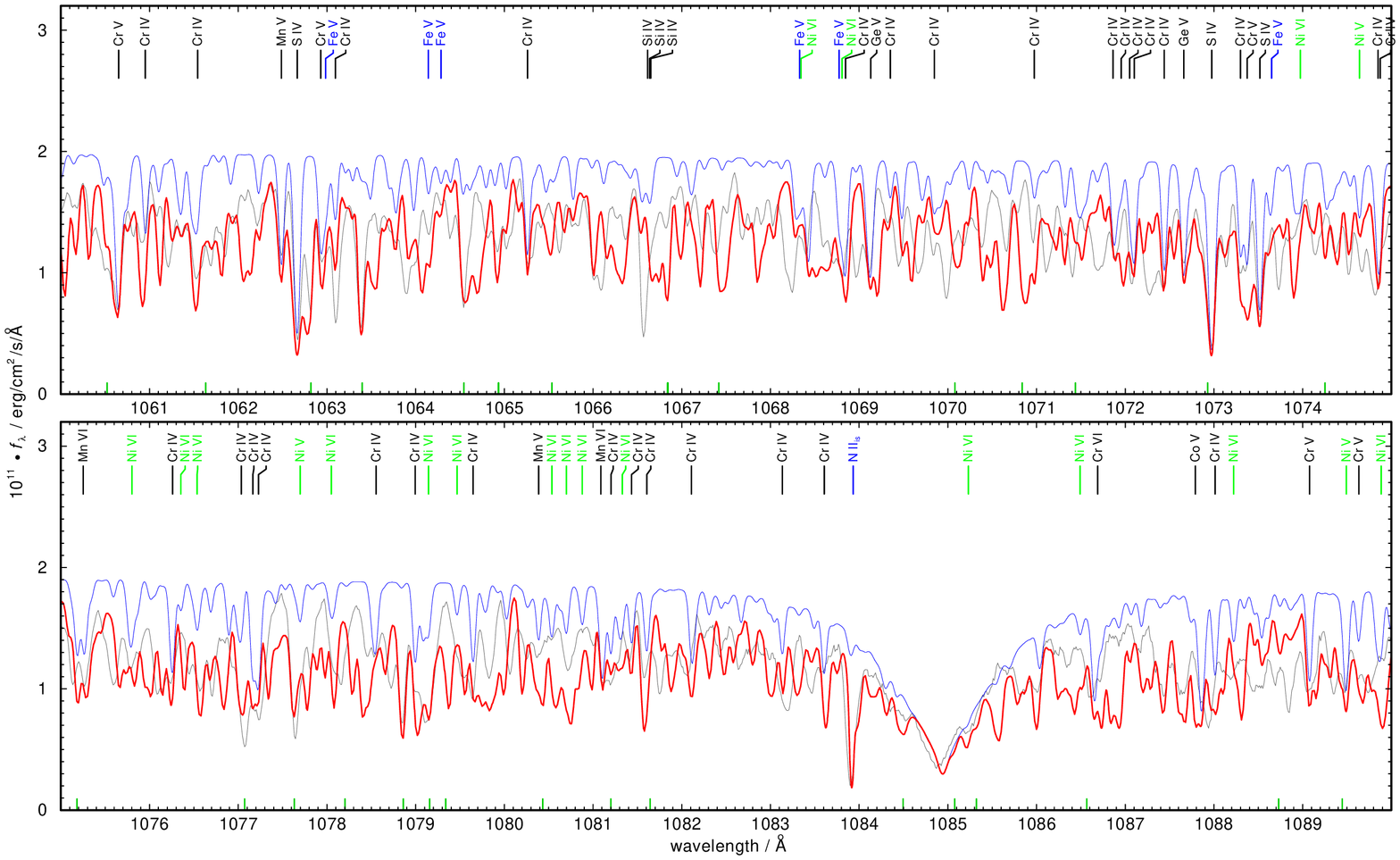}
  \caption{Continued.}
\end{figure*}
\end{landscape}
}

\addtocounter{figure}{-1}

\onlfig{
\begin{landscape}
\addtolength{\textwidth}{6.3cm} 
\addtolength{\evensidemargin}{1cm}
\addtolength{\oddsidemargin}{1cm}
\begin{figure*}
\includegraphics[trim=0 0 0 0,height=24.5cm,angle=270]{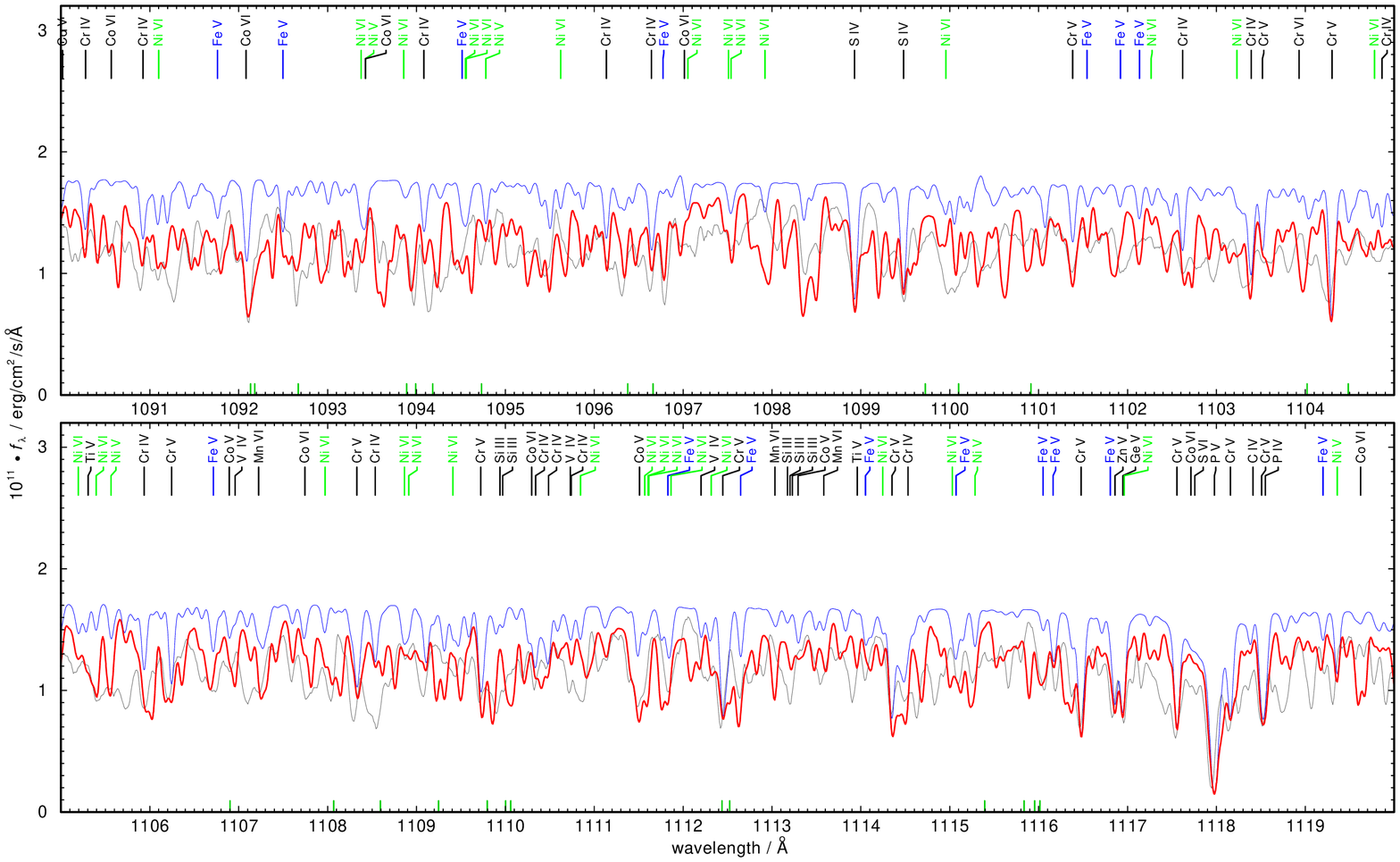}
  \caption{Continued.}
\end{figure*}
\end{landscape}
}

\addtocounter{figure}{-1}

\onlfig{
\begin{landscape}
\addtolength{\textwidth}{6.3cm} 
\addtolength{\evensidemargin}{1cm}
\addtolength{\oddsidemargin}{1cm}
\begin{figure*}
\includegraphics[trim=0 0 0 0,height=24.5cm,angle=270]{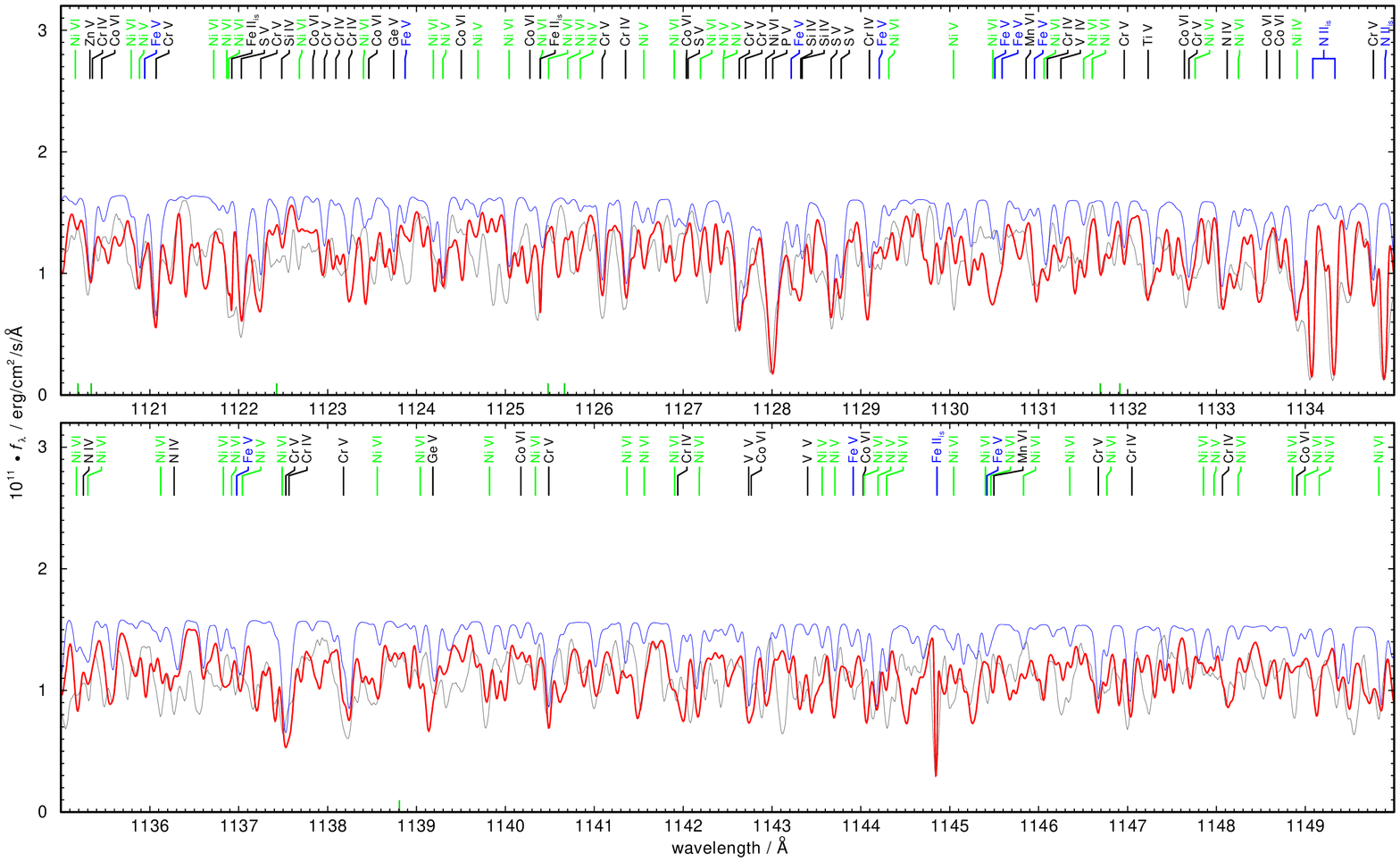}
  \caption{Continued.}
\end{figure*}
\end{landscape}
}

\addtocounter{figure}{-1}

\onlfig{
\begin{landscape}
\addtolength{\textwidth}{6.3cm} 
\addtolength{\evensidemargin}{1cm}
\addtolength{\oddsidemargin}{1cm}
\begin{figure*}
\includegraphics[trim=0 0 0 0,height=24.5cm,angle=270]{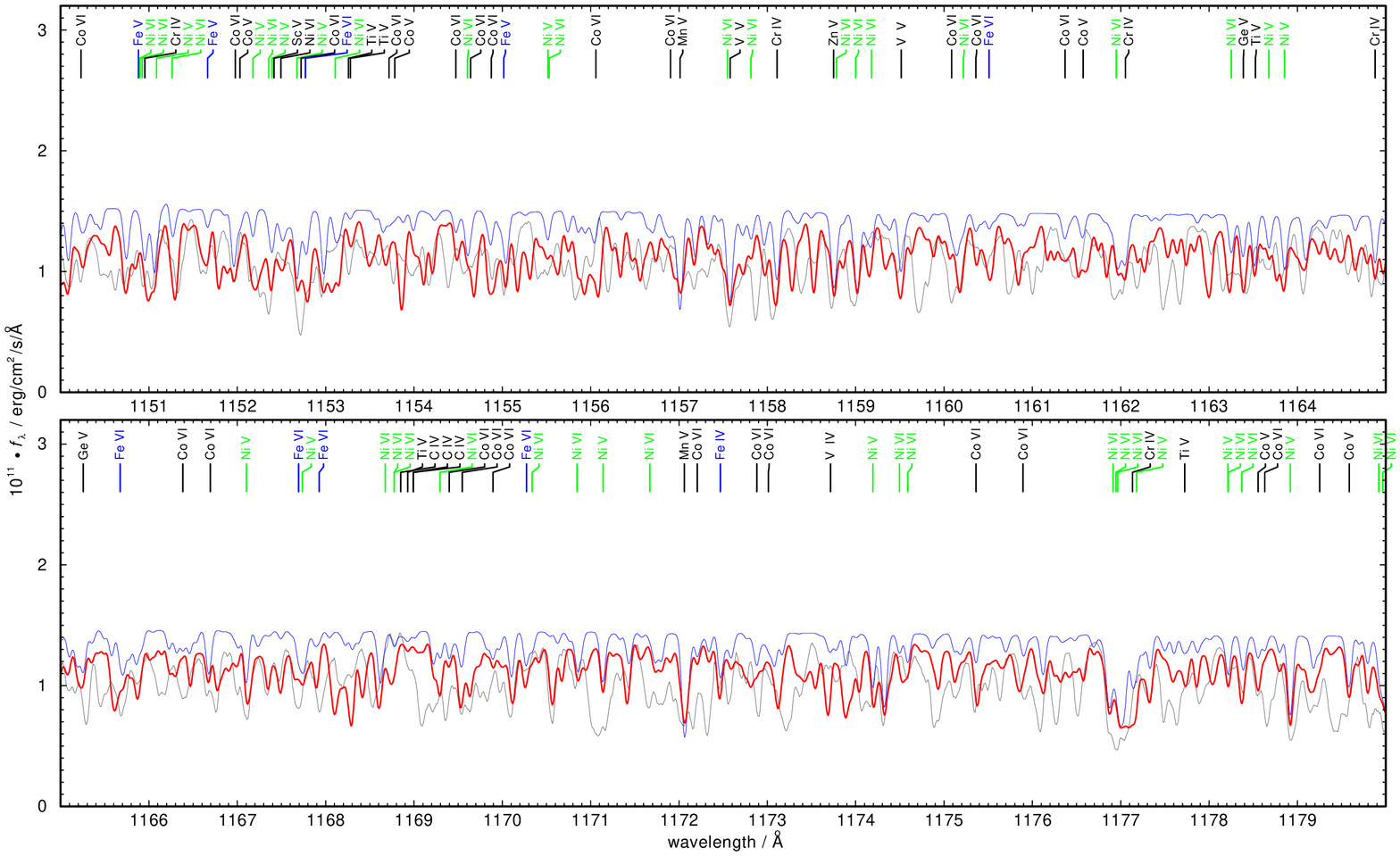}
  \caption{Continued.}
\end{figure*}
\end{landscape}
}

\addtocounter{figure}{-1}

\onlfig{
\begin{landscape}
\addtolength{\textwidth}{6.3cm} 
\addtolength{\evensidemargin}{1cm}
\addtolength{\oddsidemargin}{1cm}
\begin{figure*}
\includegraphics[trim=0 0 0 0,height=24.5cm,angle=270]{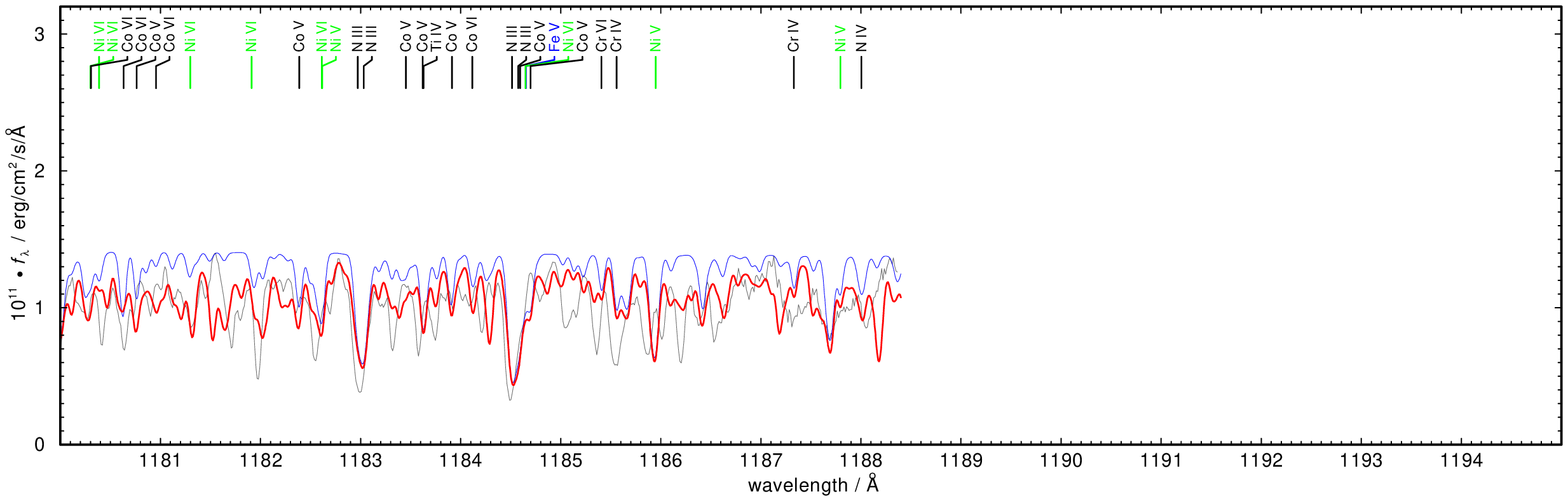}
  \caption{Continued.}
\end{figure*}
\end{landscape}
}

\section{Results and conclusions}
\label{sect:results}

\begin{figure}
  \resizebox{\hsize}{!}{\includegraphics{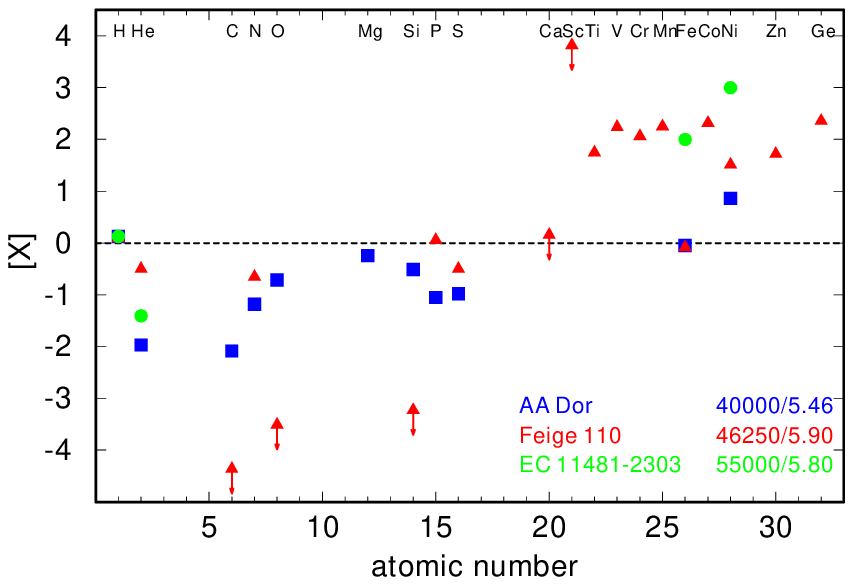}}
  \caption{Comparison of the determined photospheric abundances (arrows indicate upper limits)
           of the three OB-type subdwarfs 
           \object{AA\,Dor} \citep{klepprauch2011}, 
           \object{EC\,11481$-$2303} \citet{rauchetal2010b}, and 
           \fg. 
           Their \Teff and \logg are shown in the legend.} 
  \label{fig:abund}
\end{figure}

We performed a comprehensive spectral analysis of \fg, based on observations from the
FUV to the optical wavelength range. 
We determined
\Teffw{47\,250 \pm 2000} and 
\loggw{6.00 \pm 0.20}. 
The ionization equilibria of
\ion{He}{i}  / \ion{He}{ii},
\ion{N}{iii} / \ion{N}{iv} / \ion{N}{v},
\ion{P}{iv}  / \ion{P}{v},
\ion{S}{iv}  / \ion{S}{v}  / \ion{S}{vi},
\ion{Ti}{iv} / \ion{Ti}{v}, 
\ion{V}{iv}  / \ion{V}{v}, 
\ion{Cr}{iv} / \ion{Cr}{v} / \ion{Cr}{vi}, 
\ion{Mn}{v}  / \ion{Mn}{vi}, 
\ion{Fe}{v}  / \ion{Fe}{vi}, 
\ion{Co}{v}  / \ion{Co}{vi}, and
\ion{Ni}{v}  / \ion{Ni}{vi} 
are well reproduced with these values.
The photospheric abundances were determined based on the FUSE and optical observations (Table\,\ref{tab:abund}).
Figure\,\ref{fig:abund} shows a comparison of the photospheric abundances patterns
of three hot O(B)-type subdwarfs. While the intermediate-mass metals are solar or
subsolar in all these stars, the iron-group elements but Fe have strongly super-solar values.
An exception is Fe in \object{AA\,Dor} and \object{EC\,11481$-$2303}, that appears to be solar.
Neither this Fe peculiarity nor the extremely low C and Si abundances in \fg
can be explained.

\begin{figure}
  \resizebox{\hsize}{!}{\includegraphics{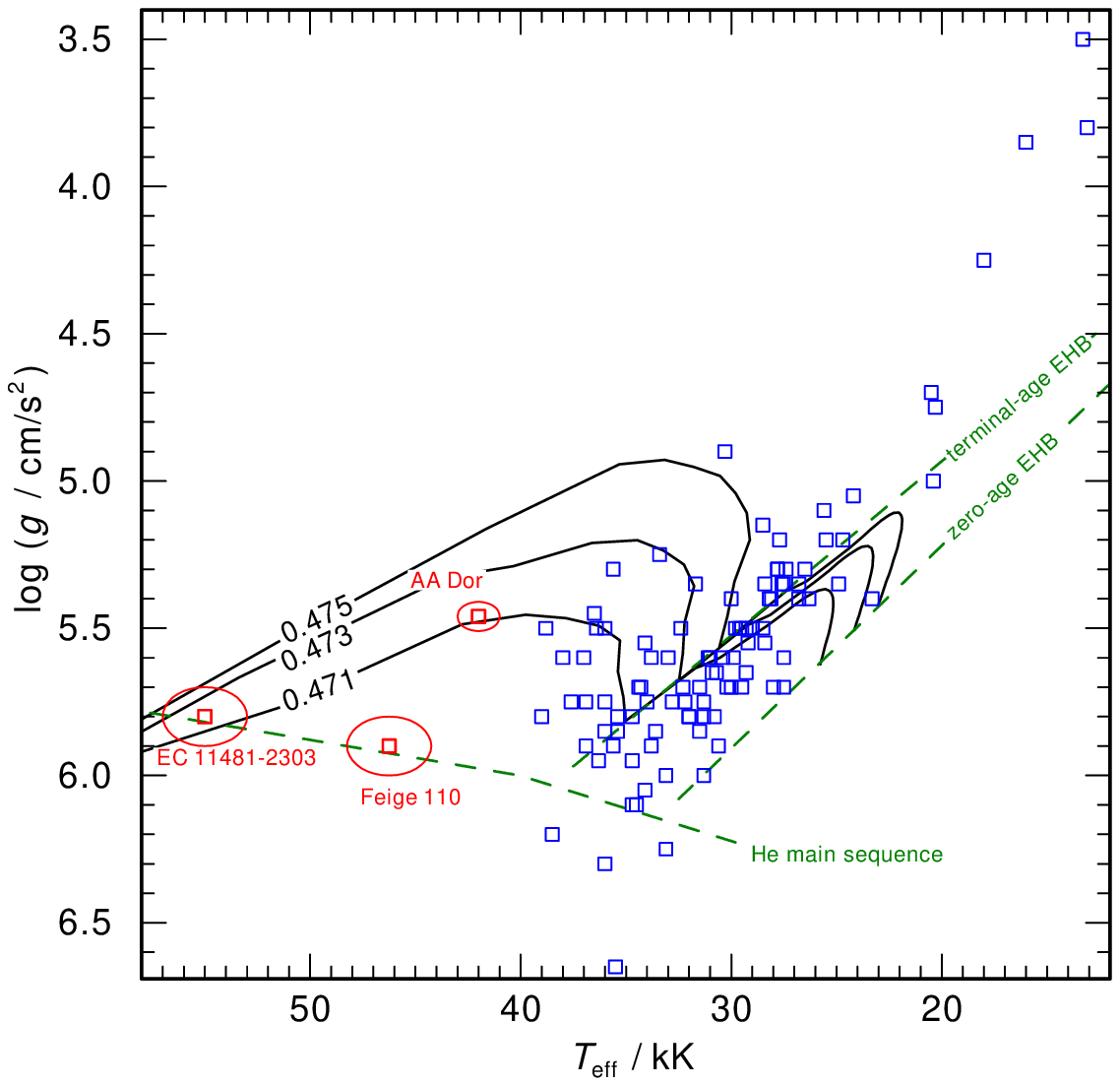}}
  \caption{Location of \fg in the \Teff $-$ \logg
           plane compared to sdBs and sdOBs from \citet{edelmann2003},
           \citet[\object{EC\,11481$-$2303}]{rauchetal2010b},
            and \citet[\object{AA\,Dor}]{klepprauch2011}.
           Post-EHB tracks from \citet[][$Y_\mathrm{HB} = 0.288$, labeled with the
           respective stellar masses in $M_\odot$]{dormanetal1993} 
           are also shown. Their start and kink points are used to
           illustrate the location of the zero-age and terminal age EHB
           for this He composition. The He main sequence is taken from
           \citet{paczynski1971}.
} 
  \label{fig:tefflogg}
\end{figure}

The position of \fg in the \Teff $-$ \logg plane shows that it is located
directly on the He main sequence (Fig.\,\ref{fig:tefflogg}). \fg belongs, 
like \object{AA\,Dor} or \object{EC\,11481$-$2303}, to the hottest post-EHB\footnote{extended
horizontal branch} stars. From a comparison to post-EHB tracks \citep{dormanetal1993}, we can 
extrapolate a stellar mass of $M = 0.469 \pm 0.001$\,\Msol. 
With $R = \sqrt{GM/g}$ ($G$ is the gravitational
constant), we calculated the stellar radius of $R = 0.114^{+0.030}_{-0.024}\,R_\odot$.

We determined the distance of \fg following the flux calibration 
of \citet{heberetal1984a} for $\lambda_\mathrm{eff} = 5454\,\mathrm{\AA}$,

\begin{equation}
d_\mathrm{spec} = 7.11 \times 10^4 \cdot \sqrt{H_\nu\cdot M \cdot 10^{0.4\, m_{\mathrm{V}_0}-\log g}}\,\mathrm{pc}\,,
\label{eq:distance}
\end{equation}

\noindent
with 
$m_\mathrm{V_o} = m_\mathrm{V} - 2.175 c$, 
$c = 1.475 E_\mathrm{B-V}$, and the Eddington flux 
$H_\nu = 7.24 \pm 0.37 \times 10^{-4}\, \mathrm{erg/cm^{2}/s/Hz}$ at $\lambda_\mathrm{eff}$ 
of our final model atmosphere.
We used 
$\ebv =0.027 \pm 0.007$ (Sect.\,\ref{sect:prelim}), 
$M = 0.469 \pm +0.001$\,\Msol, and
$m_\mathrm{V} = 11.847 \pm 0.010$ \citep{kharchenkoroeser2009}
and derived a distance of 
$d_\mathrm{spec}=297^{+62}_{-77}$\,pc 
and a height below the 
Galactic plane of 
$z=255^{+53}_{-66}$\,pc. 
This distance is about a factor of three larger than 
the new Hipparcos parallax-measurement reduction
\citep[\object{HIP115195}, $\pi = 9.76 \pm 3.44$\,mas]{vanleeuwen2007} of $d_\mathrm{parallax}=102.46^{+55.78}_{-26.69}$\,pc.
Interestingly, the older Hipparcos measurement published
by \citet[$\pi = 5.59 \pm 3.34$\,mas, $d_\mathrm{parallax}=178.89^{+265.55}_{-66.91}$\,pc]{perrymanetal1997} 
deviates from this new value by a factor of almost two and would be in agreement with our spectroscopic
distance within error limits.

The discrepancy between spectroscopic and parallax distances is a significant problem.
\logg cannot be higher by about 0.5 to achieve a distance agreement, because the spectral lines in
the models appear too broad and too shallow. This apparently is not a problem of our TMAP code, because
\citet[$d_\mathrm{spec}=288 \pm 43$\,pc]{friedmanetal2002} used the TLUSTY code and encountered the same problem.
Similar discrepancies are reported by 
\citet[][for \object{LSV+46$\degr$21} with TMAP:    $d_\mathrm{spec}     = 224^{+46}_{-58}\,\mathrm{pc}$ vs\@. 
                                                    $d_\mathrm{parallax} = 129^{+6}_{-5}\,\mathrm{pc}$]{rauchetal2007}
and by
\citet[][for \object{BD+28$\degr$4211} with TLUSTY, \Teffw{82\,000}, \loggw{6.2}, and an assumed $M = 0.5$\,\Msol:
                                                    $d_\mathrm{spec}     = 157\,\mathrm{pc}$ (no error estimate given)
                                                    vs\@. 
                                                    $d_\mathrm{parallax} =  92^{+13}_{-11}\,\mathrm{pc}$]{latouretal2013}.

\citet{latouretal2013} mentioned that a relatively high \logg value and/or a low mass may be the
solution and since they regard the HIPPARCOS measurement as fully reliable and their TLUSTY results reasonably
reliable, the mass of \object{BD+28$\degr$4211} must be much less than the canonical post-EHB mass
of about 0.5\,\Msol. For their $d_\mathrm{parallax}/d_\mathrm{spec} = 0.59$, the mass has to be
about 0.17\,\Msol. In case of \fg, with $d_\mathrm{parallax}/d_\mathrm{spec} = 0.31$,
the mass has to be about 0.10\,\Msol. In both cases, the mass can be higher, if \logg is higher.
Thus, since \logg is also the main error source in the spectroscopic distance (Eq.\,\ref{eq:distance}), 
one might speculate about the applied broadening theory for lines that are used to
determine \logg. 
For the relevant \ion{H}{i} and \ion{He}{ii} lines (linear Stark effect), 
TMAP as well as TLUSTY use the same data of \citet{tremblaybergeron2009} and \citet{schoeningbutler1989}, respectively.
However, all the narrow metal lines (e.g\@. of the iron-group element) in the UV, that are broadened by the 
quadratic Stark effect, cannot be reproduced at a much higher \logg.
To summarize, the distance discrepancy is as yet unexplained.

The analysis of the FUV spectrum has shown that the lack of reliably measured wavelengths of lines
of the iron-group elements (Ca - Ni) and of elements heavier than Ni hampers the line-identification. 
Efforts in this field in the near future are highly desirable.

The established database of spectrophotometric standard stars in TheoSSA was
extended by the OB-type subdwarf \fg.
The successfully launched GAIA\footnote{\url{http://www.esa.int/Our_Activities/Space_Science/Gaia }} mission 
will provide accurate parallax measurements for spectrophotometric standard stars. This will 
strengthen the importance of a VO-compliant database like TheoSSA that provides easy access to the best 
synthetic spectra calculated for these stars.

\begin{acknowledgements}
TR is supported by the German Aerospace Center (DLR, grant 05\,OR\,1301).  
The GAVO project at T\"ubingen has been supported by the Federal Ministry of Education and
Research (BMBF, grants  05\,AC\,6\,VTB, 05\,AC\,11\,VTB).
This work used the profile-fitting procedure OWENS developed by M\@. Lemoine and the FUSE French Team.
This research has made use of the SIMBAD database, operated at CDS, Strasbourg, France.
This research has made use of NASA's Astrophysics Data System.
Some of the data presented in this paper were obtained from the 
Mikulski Archive for Space Telescopes (MAST). STScI is operated by the 
Association of Universities for Research in Astronomy, Inc., under NASA 
contract NAS5-26555. Support for MAST for non-HST data is provided by 
the NASA Office of Space Science via grant NNX09AF08G and by other 
grants and contracts.
The TIRO service (\url{http://astro-uni-tuebingen.de/~TIRO}) used to calculate
opacities for this paper was constructed as part of the
activities of the German Astrophysical Virtual Observatory.
The TMAW service (\url{http://astro-uni-tuebingen.de/~TMAW}) used to calculate
theoretical spectra for this paper was constructed as part of the
activities of the German Astrophysical Virtual Observatory.
\end{acknowledgements}

\bibliographystyle{aa}
\bibliography{23711}

\begin{thebibliography}{43}
\expandafter\ifx\csname natexlab\endcsname\relax\def\natexlab#1{#1}\fi

\bibitem[{{Asplund} {et~al.}(2009){Asplund}, {Grevesse}, {Sauval}, \&
  {Scott}}]{asplundetal2009}
{Asplund}, M., {Grevesse}, N., {Sauval}, A.~J., \& {Scott}, P. 2009, \araa, 47,
  481

\bibitem[{{Bohlin} {et~al.}(1990){Bohlin}, {Harris}, {Holm}, \&
  {Gry}}]{bohlinetal1990}
{Bohlin}, R.~C., {Harris}, A.~W., {Holm}, A.~V., \& {Gry}, C. 1990, \apjs, 73,
  413

\bibitem[{{Dorman} {et~al.}(1993){Dorman}, {Rood}, \&
  {O'Connell}}]{dormanetal1993}
{Dorman}, B., {Rood}, R.~T., \& {O'Connell}, R.~W. 1993, \apj, 419, 596

\bibitem[{{Edelmann}(2003)}]{edelmann2003}
{Edelmann}, H. 2003, Dissertation, University Erlangen-Nuremberg, Germany

\bibitem[{{Friedman} {et~al.}(2002){Friedman}, {Howk}, {Chayer}, {Tripp},
  {H{\'e}brard}, {Andr{\'e}}, {Oliveira}, {Jenkins}, {Moos}, {Oegerle},
  {Sonneborn}, {Lamontagne}, {Sembach}, \& {Vidal-Madjar}}]{friedmanetal2002}
{Friedman}, S.~D., {Howk}, J.~C., {Chayer}, P., {et~al.} 2002, \apjs, 140, 37

\bibitem[{{Good} {et~al.}(2004){Good}, {Barstow}, {Holberg}, {Sing},
  {Burleigh}, \& {Dobbie}}]{goodetal2004}
{Good}, S.~A., {Barstow}, M.~A., {Holberg}, J.~B., {et~al.} 2004, \mnras, 355,
  1031

\bibitem[{{Heber} {et~al.}(1984{\natexlab{a}}){Heber}, {Hamann}, {Hunger},
  {Kudritzki}, {Simon}, \& {Mendez}}]{heberetal1984b}
{Heber}, U., {Hamann}, W.-R., {Hunger}, K., {et~al.} 1984{\natexlab{a}}, \aap,
  136, 331

\bibitem[{{Heber} {et~al.}(1984{\natexlab{b}}){Heber}, {Hunger}, {Jonas}, \&
  {Kudritzki}}]{heberetal1984a}
{Heber}, U., {Hunger}, K., {Jonas}, G., \& {Kudritzki}, R.~P.
  1984{\natexlab{b}}, \aap, 130, 119

\bibitem[{{H{\'e}brard} {et~al.}(2002){H{\'e}brard}, {Friedman}, {Kruk},
  {Lehner}, {Lemoine}, {Linsky}, {Moos}, {Oliveira}, {Sembach}, {Sonneborn},
  {Vidal-Madjar}, \& {Wood}}]{hebrard02}
{H{\'e}brard}, G., {Friedman}, S.~D., {Kruk}, J.~W., {et~al.} 2002, \planss,
  50, 1169

\bibitem[{{H{\'e}brard} \& {Moos}(2003)}]{hebrard03}
{H{\'e}brard}, G. \& {Moos}, H.~W. 2003, \apj, 599, 297

\bibitem[{{Hubeny} \& {Lanz}(1995)}]{hubenylanz1995}
{Hubeny}, I. \& {Lanz}, T. 1995, \apj, 439, 875

\bibitem[{{Kharchenko} \& {Roeser}(2009)}]{kharchenkoroeser2009}
{Kharchenko}, N.~V. \& {Roeser}, S. 2009, VizieR Online Data Catalog, 1280, 0

\bibitem[{{Klepp} \& {Rauch}(2011)}]{klepprauch2011}
{Klepp}, S. \& {Rauch}, T. 2011, \aap, 531, L7

\bibitem[{{Kudritzki}(1976)}]{kudritzki1976}
{Kudritzki}, R.-P. 1976, \aap, 52, 11

\bibitem[{{Kurucz}(1991)}]{kurucz1991}
{Kurucz}, R.~L. 1991, in NATO ASIC Proc. 341: Stellar Atmospheres - Beyond
  Classical Models, ed. L.~{Crivellari}, I.~{Hubeny}, \& D.~G. {Hummer}, 441

\bibitem[{{Kurucz}(2009)}]{kurucz2009}
{Kurucz}, R.~L. 2009, in American Institute of Physics Conference Series, Vol.
  1171, American Institute of Physics Conference Series, ed. I.~{Hubeny}, J.~M.
  {Stone}, K.~{MacGregor}, \& K.~{Werner}, 43

\bibitem[{{Kurucz}(2011)}]{kurucz2011}
{Kurucz}, R.~L. 2011, Canadian Journal of Physics, 89, 417

\bibitem[{{Latour} {et~al.}(2013){Latour}, {Fontaine}, {Chayer}, \&
  {Brassard}}]{latouretal2013}
{Latour}, M., {Fontaine}, G., {Chayer}, P., \& {Brassard}, P. 2013, \apj, 773,
  84

\bibitem[{{Liszt}(2014{\natexlab{a}})}]{liszt2014a}
{Liszt}, H. 2014{\natexlab{a}}, \apj, 783, 17

\bibitem[{{Liszt}(2014{\natexlab{b}})}]{liszt2014b}
{Liszt}, H. 2014{\natexlab{b}}, \apj, 780, 10

\bibitem[{{Moehler} {et~al.}(2014){Moehler}, {Modigliani}, {Freudling},
  {Giammichele}, {Gianninas}, {Gonneau}, {Kausch}, {Koester}, {Lan\c{c}on},
  {Noll}, {Rauch}, \& {Vinther}}]{moehleretal2014}
{Moehler}, S., {Modigliani}, A., {Freudling}, W., {et~al.} 2014, \aap, in press

\bibitem[{{Napiwotzki} \& {Rauch}(1994)}]{napiwotzkirauch1994}
{Napiwotzki}, R. \& {Rauch}, T. 1994, \aap, 285, 603

\bibitem[{{Newell}(1973)}]{newell1973}
{Newell}, E.~B. 1973, \apjs, 26, 37

\bibitem[{{Oke}(1990)}]{oke1990}
{Oke}, J.~B. 1990, \aj, 99, 1621

\bibitem[{{Paczy{\'n}ski}(1971)}]{paczynski1971}
{Paczy{\'n}ski}, B. 1971, \actaa, 21, 1

\bibitem[{{Perryman} {et~al.}(1997){Perryman}, {Lindegren}, {Kovalevsky},
  {Hoeg}, {Bastian}, {Bernacca}, {Cr{\'e}z{\'e}}, {Donati}, {Grenon},
  {Grewing}, {van Leeuwen}, {van der Marel}, {Mignard}, {Murray}, {Le Poole},
  {Schrijver}, {Turon}, {Arenou}, {Froeschl{\'e}}, \&
  {Petersen}}]{perrymanetal1997}
{Perryman}, M.~A.~C., {Lindegren}, L., {Kovalevsky}, J., {et~al.} 1997, \aap,
  323, L49

\bibitem[{{Rauch}(1997)}]{rauch1997}
{Rauch}, T. 1997, \aap, 320, 237

\bibitem[{{Rauch}(2000)}]{rauch2000}
{Rauch}, T. 2000, \aap, 356, 665

\bibitem[{{Rauch}(2003)}]{rauch2003}
{Rauch}, T. 2003, \aap, 403, 709

\bibitem[{{Rauch} \& {Deetjen}(2003)}]{rauchdeetjen2003}
{Rauch}, T. \& {Deetjen}, J.~L. 2003, in Astronomical Society of the Pacific
  Conference Series, Vol. 288, Stellar Atmosphere Modeling, ed. I.~{Hubeny},
  D.~{Mihalas}, \& K.~{Werner}, 103

\bibitem[{{Rauch} {et~al.}(1998){Rauch}, {Dreizler}, \&
  {Wolff}}]{rauchetal1998}
{Rauch}, T., {Dreizler}, S., \& {Wolff}, B. 1998, \aap, 338, 651

\bibitem[{{Rauch} {et~al.}(2013){Rauch}, {Werner}, {Bohlin}, \&
  {Kruk}}]{rauchetal2013}
{Rauch}, T., {Werner}, K., {Bohlin}, R., \& {Kruk}, J.~W. 2013, \aap, 560, A106

\bibitem[{{Rauch} {et~al.}(2010){Rauch}, {Werner}, \& {Kruk}}]{rauchetal2010b}
{Rauch}, T., {Werner}, K., \& {Kruk}, J.~W. 2010, \apss, 329, 133

\bibitem[{{Rauch} {et~al.}(2007){Rauch}, {Ziegler}, {Werner}, {Kruk},
  {Oliveira}, {Vande Putte}, {Mignani}, \& {Kerber}}]{rauchetal2007}
{Rauch}, T., {Ziegler}, M., {Werner}, K., {et~al.} 2007, \aap, 470, 317

\bibitem[{{Sch\"oning} \& {Butler}(1989)}]{schoeningbutler1989}
{Sch\"oning}, T. \& {Butler}, K. 1989, \aaps, 78, 51

\bibitem[{{Tremblay} \& {Bergeron}(2009)}]{tremblaybergeron2009}
{Tremblay}, P.-E. \& {Bergeron}, P. 2009, \apj, 696, 1755

\bibitem[{{Turnshek} {et~al.}(1990){Turnshek}, {Bohlin}, {Williamson}, {Lupie},
  {Koornneef}, \& {Morgan}}]{turnsheketal1990}
{Turnshek}, D.~A., {Bohlin}, R.~C., {Williamson}, II, R.~L., {et~al.} 1990,
  \aj, 99, 1243

\bibitem[{{van Leeuwen}(2007)}]{vanleeuwen2007}
{van Leeuwen}, F. 2007, \aap, 474, 653

\bibitem[{{Vennes} {et~al.}(2011){Vennes}, {Kawka}, \&
  {N{\'e}meth}}]{vennesetal2011}
{Vennes}, S., {Kawka}, A., \& {N{\'e}meth}, P. 2011, \mnras, 410, 2095

\bibitem[{{Vernet} {et~al.}(2011){Vernet}, {Dekker}, {D'Odorico}, {Kaper},
  {Kjaergaard}, {Hammer}, {Randich}, {Zerbi}, {Groot}, {Hjorth}, {Guinouard},
  {Navarro}, {Adolfse}, {Albers}, {Amans}, {Andersen}, {Andersen}, {Binetruy},
  {Bristow}, {Castillo}, {Chemla}, {Christensen}, {Conconi}, {Conzelmann},
  {Dam}, {de Caprio}, {de Ugarte Postigo}, {Delabre}, {di Marcantonio},
  {Downing}, {Elswijk}, {Finger}, {Fischer}, {Flores}, {Fran{\c c}ois},
  {Goldoni}, {Guglielmi}, {Haigron}, {Hanenburg}, {Hendriks}, {Horrobin},
  {Horville}, {Jessen}, {Kerber}, {Kern}, {Kiekebusch}, {Kleszcz}, {Klougart},
  {Kragt}, {Larsen}, {Lizon}, {Lucuix}, {Mainieri}, {Manuputy}, {Martayan},
  {Mason}, {Mazzoleni}, {Michaelsen}, {Modigliani}, {Moehler}, {M{\o}ller},
  {Norup S{\o}rensen}, {N{\o}rregaard}, {P{\'e}roux}, {Patat}, {Pena}, {Pragt},
  {Reinero}, {Rigal}, {Riva}, {Roelfsema}, {Royer}, {Sacco}, {Santin},
  {Schoenmaker}, {Spano}, {Sweers}, {Ter Horst}, {Tintori}, {Tromp}, {van
  Dael}, {van der Vliet}, {Venema}, {Vidali}, {Vinther}, {Vola}, {Winters},
  {Wistisen}, {Wulterkens}, \& {Zacchei}}]{vernetetal2011}
{Vernet}, J., {Dekker}, H., {D'Odorico}, S., {et~al.} 2011, \aap, 536, A105

\bibitem[{{Wassermann} {et~al.}(2010){Wassermann}, {Werner}, {Rauch}, \&
  {Kruk}}]{wassermannetal2010}
{Wassermann}, D., {Werner}, K., {Rauch}, T., \& {Kruk}, J.~W. 2010, \aap, 524,
  A9

\bibitem[{{Werner} {et~al.}(2003){Werner}, {Deetjen}, {Dreizler}, {Nagel},
  {Rauch}, \& {Schuh}}]{werneretal2003}
{Werner}, K., {Deetjen}, J.~L., {Dreizler}, S., {et~al.} 2003, in Astronomical
  Society of the Pacific Conference Series, Vol. 288, Stellar Atmosphere
  Modeling, ed. I.~{Hubeny}, D.~{Mihalas}, \& K.~{Werner}, 31

\bibitem[{{Ziegler} {et~al.}(2012){Ziegler}, {Rauch}, {Werner}, {K{\"o}ppen},
  \& {Kruk}}]{ziegleretal2012}
{Ziegler}, M., {Rauch}, T., {Werner}, K., {K{\"o}ppen}, J., \& {Kruk}, J.~W.
  2012, \aap, 548, A109

\end{thebibliography}

\end{document}